\documentstyle[prb,aps]{revtex}
\begin{document}
\draft
\title{Surface acoustic wave attenuation by a two-dimensional electron gas
       in a strong magnetic field}
\author{Andreas Kn\"abchen and
	Yehoshua B.\ Levinson}
\address{Weizmann Institute of Science,
Department of Condensed Matter Physics,\\
76100 Rehovot, Israel}
\author{Ora Entin-Wohlman}
\address{School of Physics and Astronomy,
Raymond and Beverly Sackler Faculty of Exact Sciences,\\
Tel Aviv University, 69978 Tel Aviv, Israel}
\date{\today}
\maketitle
\begin{abstract}
The propagation of a surface acoustic wave (SAW)
on GaAs/AlGaAs heterostructures is studied in the case where
the two-dimensional electron gas (2DEG) is subject to a strong magnetic field
and a smooth random potential with correlation length $\Lambda$
and amplitude $\Delta$.
The electron wave functions are described in a quasiclassical
picture using results of percolation theory for
two-dimensional systems.
In accordance with the experimental situation, $\Lambda$ is assumed to be much smaller
than the sound wavelength $2\pi/q$. This restricts 
the absorption of surface phonons
at a filling factor $\bar{\nu} \approx 1/2$ to
electrons occupying extended trajectories of fractal structure.
Both piezoelectric and deformation potential interactions of
surface acoustic phonons with electrons are considered and
the corresponding interaction vertices are derived.
These vertices are found to differ from those valid for three-dimensional bulk
phonon systems with respect to the phonon wave vector dependence.
We derive the appropriate dielectric function 
$\varepsilon(\omega,q)$ to describe the effect of screening on the
electron-phonon coupling. 
In the low temperature, high frequency regime
$T\ll\Delta (\omega_q\Lambda /v_D)^{\alpha/2\nu}$,
where $\omega_q$ is the SAW frequency
and $v_D$ is the electron drift velocity,
both the attenuation coefficient $\Gamma$ and 
$\varepsilon(\omega,q)$ are independent of temperature.
The classical percolation indices give $\alpha/2\nu=3/7$.
The width of the region where a strong absorption of the SAW occurs
is found to be given by the scaling law
$|\Delta \bar{\nu}| \approx (\omega_q \Lambda /v_D)^{\alpha/2\nu}$.
The dependence of the electron-phonon coupling and the screening due to
the 2DEG on the filling factor leads to a double-peak structure for
$\Gamma(\bar\nu)$. 
\end{abstract}

\section{Introduction}\label{intro}

Surface acoustic waves (SAW's) \cite{Farnell78,Mayer95} provide a useful tool
for experimental studies of the two-dimensional electron gas (2DEG)
in GaAs/AlGaAs-heterostructures.
In particular, SAW's have been used in recent years in investigations of the integer
\cite{Rampton92,Wixforth86,Wixforth89,Schenstrom88,Esslinger94} and the
fractional \cite{Esslinger94,Willet90,Willet94}
quantum Hall regimes.
Due to the quantum Hall effect, the interaction of the SAW with the charge
carriers can lead to strong oscillations in the
attenuation and the velocity of the sound waves as function
of the applied magnetic field.
Quantum oscillations have also been reported
for the sound-induced currents and voltages \cite{Esslinger94,Shilton95}.

Previous theoretical descriptions of these experiments have been based essentially
on classical models for the propagation of SAW's \cite{McFee66,Tucker72}.
According to these models, which are originally derived for systems
in the absence of an applied magnetic field, the sound attenuation is expressed
in terms of the electrical dc conductivity. This relation is derived under the
assumptions that
$ql \ll 1$ and \mbox{$\omega_q \tau \ll 1$} (`local regime'), where \mbox{\boldmath $q$}
and $\omega_q$
are the wave vector and the frequency of the sound, respectively, and
$l$ and $\tau$ are the mean free path and the scattering time of the
conduction electrons, respectively.
If a (classical) magnetic field is applied, the first condition has to be
replaced
by $qR_c \ll 1$, where $R_c$ is the cyclotron radius\cite{Tucker72}.
It is much more difficult, however, to determine under which conditions the 
above mentioned
theories are valid when the 2DEG is subject to a quantizing magnetic field.
In this case, the electron system is characterized by (at least)
two more length scales, namely the magnetic length $l_B=\sqrt{c\hbar/eB}$
and the localization length \cite{Jansen94} $\xi$.
While $ql_B \ll 1$ is always fulfilled under typical experimental
conditions, the localization length can be much larger than
the surface acoustic wavelength $2\pi/q$.

A series of experiments
\cite{Rampton92,Wixforth86,Wixforth89,Schenstrom88,Esslinger94,Willet90,Willet94}
has shown a reasonable agreement with the predictions of the classical
models in a wide range of frequencies $\omega_q$ and magnetic field strengths.
On the other hand, some deviations have also been detected.
For example, deviations of the
SAW attenuation from the classically predicted behavior with increasing
frequency have been reported\cite{Wixforth89}. These
were attributed to nonlocal effects of the interaction between
the SAW and the 2DEG which should occur when the sound wavelength becomes
of the order of or smaller
than a characteristic length scale of the electron gas.
In the fractional quantum Hall regime,
an anomaly in the absorption coefficient for filling factor $\bar{\nu}=1/2$ was
found \cite{Willet90,Willet94}.
This anomaly was discussed in the framework of the composite Fermion model
of Ref.\ \onlinecite{Halperin93}. According to this approach, the electrons
are replaced by composite Fermions moving in an effective magnetic field of zero
average (at $\bar{\nu}=1/2$).
Then, the sound absorption due to these particles is described
by classical formulae except that the dc conductivity is replaced by the wave
vector
dependent nonlocal conductivity which, however,
represents a very important difference
(see Ref.\ \onlinecite{Halperin93} for details).

In this paper  we study the propagation of SAW's in
the integer quantum Hall regime. The calculation of the SAW attenuation
is carried out for filling factors near 1/2 and is based
on a percolation approach to the electronic states in a very strong magnetic
field.
From this point of view
we may anticipate a nonlocal behavior of the attenuation arising from
the large characteristic length scales
(e.~g., the size of a percolation cluster $\gg q^{-1}$)
inherent in that framework. The effect of electron-electron
interaction is taken into account by the screening
of the electron-phonon  coupling.
The same problem has also been studied
in Ref.\ \onlinecite{Aleiner94}.
These authors calculated the nonlocal conductivity due to
variable-range hopping between pairs of localized states. Then, in the
spirit of the classical description of sound absorption,
this conductivity is related to the SAW attenuation coefficient.
A comparison with our results will be given in Sec.\ VI.

The system considered is
a 2DEG, subject to a very strong magnetic field $B$
and a smooth random potential $V$.
The potential can be characterized by its amplitude $\Delta$ and its
correlation length $\Lambda$. The amplitude also determines the width
of the Landau levels. The correlation length is of the order
of the spacer layer that separates the 2DEG from the dopant layers.
Under the assumption that $l_B \ll \Lambda$
a quasiclassical description of the electron states can be applied.
That is, one considers the drift motion of the guiding center of an
electron on the equipotential lines (EL's) of $V$
separately from the rapid motion relative to it \cite{Jansen94}.
The drift velocity is given by $v_D=c|\nabla V|/eB$.
Using $|\nabla V| \simeq \Delta/\Lambda$, the drift velocity can
be estimated to be $v_D \simeq l_B^2 \Delta / \hbar \Lambda$.
Depending on the ratio between the correlation length $\Lambda$ and the 
sound wavelength,
two regimes can be distinguished. For $q\Lambda \gg 1$,
the electron-phonon interaction can be considered locally neglecting
the global structure of the EL's \cite{Zhao93,Heinonen84}.
[For single-phonon absorption and emission processes to occur, one has also to require
that the sound velocity $v_s$ is smaller than the drift velocity.
This is usually referred to as $\check{\mbox{\rm C}}$erenkov condition\cite{Iordansky96}.]
This regime is valid for, e.\ g., thermal phonons\cite{Zhao93}.
However, SAW's have a much larger wavelength, and hence
$q\Lambda \ll 1$ is typically fulfilled.
In this case, 
the local absorption and emission of phonons is exponentially small, and
the EL as a whole has to be considered. It becomes important that
the motion of the guiding center on an extended EL [with a length $\gg \Lambda$]
resembles a random walk, with a diffusion coefficient $D=v_D \Lambda$.
Since the ratio $v_D/v_s$, for real systems, is not very different from
unity, one deals with the limit $\omega_q \gg Dq^2$ of this diffusion process.
[To be precise\cite{Iordansky96}, the parameter $v_D/v_s$ has to lie
in the range $q\Lambda \ll v_D/v_s \ll (q\Lambda)^{-3/4}$.]
Indeed, for
$B=10$T ($l_B=8$nm), $\Delta=1$meV,
$\Lambda\approx 50$nm, $q=10^4 $cm${}^{-1}$, and $v_s\approx 3\times 10^5$cms${}^{-1}$
we find
$q\Lambda= 0.05$ and $v_D\approx 0.7v_s$.
It is this particular (diffusive) regime
which will be addressed in this paper. 
In the same regime, the electron life-time and
the energy relaxation time due to interaction with
3D bulk phonons 
have been calculated in Ref.~\onlinecite{Iordansky96}.

The quantum mechanical calculation of the attenuation
coefficient (as well as of other quantities associated with SAW's)
requires the knowledge of the Hamiltonians which describe the interaction
of electrons with acoustic surface phonons.
As far as we know these interaction Hamiltonians have not yet been
derived.
Instead, many theoretical investigations have addressed the interaction
of 2D electrons with 3D (bulk) or 2D phonon systems.
The latter one, a single layer of vibrating
atoms, represents merely a theoretical construction. Three-dimensional phonons do not provide
an appropriate approach when the 2DEG is located near a free crystal
surface \cite{Badalyan88}. This implies that it is by no means clear that the
interaction of a SAW with the 2DEG
is described well by the formulae which are valid in the case of bulk
phonons.
In fact, we find that the interaction vertices appearing in the
general electron-phonon
interaction Hamiltonian [see Eq.\ (\ref{hdefdet})] differ from those
for 3D phonons
not only by numerical constants but also in the 
phonon wave vector dependence and the relative phase
between the deformation potential and the piezoelectric
interactions.

The paper is organized as follows.
The interaction vertices are derived and discussed in the next
section. In Sec.\ III, we describe the quasiclassical electron states
of a 2DEG in a strong magnetic field and a random potential
$V$. We show that the absorption of the SAW and the dielectric function
depend crucially on the occupation and the properties of electron states
which correspond to very long EL's.
The structure of these EL's is deduced from the
2D percolation theory. The matrix elements for transitions between
different electron states are given in Sec.\ IV.
The screening of the electron-phonon interaction due to the 2DEG
is accounted for by a dielectric
function $\varepsilon(\omega_q, q)$ which is calculated in Sec.\ V.
Based on these results, the SAW attenuation coefficient
is obtained in Sec.\ VI. Its dependence on
the filling factor (or the Fermi energy),
the SAW frequency and the temperature are discussed.
A short summary is given in Sec.\ VII.

\section{Interaction Hamiltonians}
\subsection{The displacement field}

To simplify the calculations we use the following
assumptions. Since the SAW wavelength $2\pi /q$
is much longer than the lattice constant, the crystal can be
approximated by a continuous medium. Its elastic properties
are assumed to be isotropic.
Furthermore, we disregard the fact that the GaAs-substrate is coated with layers
which differ slightly in their elastic properties. The overall thickness of these
layers
\cite{Wixforth89} $d \simeq 100$nm is
much smaller than the wavelength of sound. It has been shown \cite{Mayer95}
that for $qd \ll 1$ the deviations of the
wave propagation resulting from a thin overlayer coating
an homogeneous substrate
can be accounted for by a systematic expansion in this small parameter.
In our case $qd \le 10^{-1}$, i.~e., these corrections are indeed negligible.
Thus, we end up with the standard problem of sound waves which
are propagated in an isotropic medium bounded by a plane
\cite{Landau70,Mayer95}. (Effects resulting from the anisotropy of the lattice
become important for $qd \approx 1$, see Ref.\ \onlinecite{Simon96}.)

Let the surface be in the $x$-$y$-plane and the medium in the half space
$z \ge 0$. The longitudinal and transversal components of the displacement field
$\mbox{\boldmath $u$}(\mbox{\boldmath ${r}$},t)$, 
$\mbox{\boldmath $r$}=(x,y,z)\equiv (\mbox{\boldmath $R$},z)$, obey the wave
 equations
\begin{equation}\label{waveeq}
\frac{\partial^2 \mbox{\boldmath $u$}_{l,t}}{\partial t^2} -c_{l,t}^2 \Delta 
\mbox{\boldmath $u$}_{l,t} =0
 \, ,
\end{equation}%
where $c_{l,t}$ are the corresponding sound velocities.
By definition, {\rm curl}$\, \mbox{\boldmath $u$}_l=0$ and {\rm div}$\, 
\mbox{\boldmath $u$}_t=0$.
Surface waves are composed of particular solutions of Eqs.\ (\ref{waveeq})
that decay exponentially with increasing distance
from the surface. In addition, these solutions have to satisfy the boundary
conditions
at the free surface $z=0$, namely, the normal components of the stress tensor
should vanish there. It turns out that these boundary conditions can only be fulfilled
by a linear combination of $\mbox{\boldmath $u$}_l$ and 
$\mbox{\boldmath $u$}_t$, i.~e.,
pure longitudinal or transversal surface waves do not exist
\cite{Landau70}.
The full displacement
field for a mode with a two-dimensional wave vector \mbox{\boldmath $q$} can be written as
\begin{mathletters}\label{uq}
\begin{equation}\label{uq1}
\mbox{\boldmath $u$}_{\mbox{\boldmath $q$}} (\mbox{\boldmath $r$},t) = 
C_q e^{i(\mbox{\boldmath $qR$} -\omega_q t)}
 \mbox{\boldmath $v$}_{\mbox{\boldmath $q$}} (z)
+ {\rm c.c.} \, ,
\end{equation}%
with
\begin{equation}\label{uq2}
\mbox{\boldmath $v$}_{\mbox{\boldmath $q$}} (z) =
-i \hat{\mbox{\boldmath $q$}} (e^{-\kappa_l q z} - f \kappa_t e^{-\kappa_t qz})
+  \hat{\mbox{\boldmath $z$}} (\kappa_l e^{-\kappa_l q z} -f e^{-\kappa_t qz}).
\end{equation}%
\end{mathletters}%
That is, the displacement $\mbox{\boldmath $u$}_{\mbox{\boldmath $q$}}$ is polarized
in the sagittal plane which is spanned by the propagation direction
$\hat{\mbox{\boldmath $q$}}=\mbox{\boldmath $q$}/q$ and the surface normal 
$\hat{\mbox{\boldmath $z$}}$.
The decay of the displacements into the interior of the medium is described
by
\begin{equation}\label{kappa}
\kappa_l(\alpha) =\sqrt{1- \alpha \xi^2} \qquad {\rm and} \qquad
\kappa_t (\alpha)= \sqrt{1- \xi^2} \, ,
\end{equation}%
where $\alpha= c_t^2/c_l^2$ and $\xi$ is a root of an algebraic
equation of sixth order containing the parameter $\alpha$ only
(see p.~104 in Ref.~\onlinecite{Landau70}).
$\xi$ enters the dispersion relation of the surface waves in the form
\begin{equation}\label{omega}
\omega_q = \xi c_t q \equiv v_s q \, ,
\end{equation}
where $v_s$ is the SAW velocity.
Finally, the factor $f$ is given by

\begin{equation}\label{factor}
f(\alpha)=
\frac{1+\kappa_t^2}{2\kappa_t} = \sqrt{\frac{\kappa_l}{\kappa_t}} .
\end{equation}%

In order to quantize the displacement field (\ref{uq}),
the normalization constant $C_q$ of each individual mode has first
to be fixed in an appropriate way. That is, the energy associated with the
mode $\mbox{\boldmath $u$}_{\mbox{\boldmath $q$}}
(\mbox{\boldmath $r$},t)$ in the normalization volume has to
coincide with the energy $\hbar \omega_q$ of the corresponding phonon.
Since the wave is propagated freely along the surface, the energy is normalized
with respect to a large but finite square of area $L^2$ in the $x$-$y$-plane.
On the contrary, no such restriction is necessary with respect to the
$z$-coordinate because $\mbox{\boldmath $u$}_{\mbox{\boldmath $q$}}$ 
decays exponentially with
increasing distance from the surface. Thus, the normalization volume
can be extended from $z=0$ to $z = \infty$ under the chosen surface area.

Adding a kinetic energy term to the potential energy \cite{Landau70} associated
 with
a displacement field $\mbox{\boldmath $u$}$, the total energy can be written as
\begin{equation}\label{energy}
E(\mbox{\boldmath $u$})=
\frac{1}{2} \rho
\int \! d^3 \mbox{\boldmath $r$} \,
\left[ (\partial \mbox{\boldmath $u$} / \partial t)^2
+
(c_l^2 -2 c_t^2 ) ({\rm div} \mbox{\boldmath $u$})^2
+
2 c_t^2 \sum\limits_{i,k} (u_{ik})^2
\right] ,
\end{equation}%
where $\rho$ is the mass density of the medium and
\begin{equation}\label{strain}
u_{ik} =\frac{1}{2} \left( \frac{ \partial u_i}{\partial x_k}
+ \frac{ \partial u_k}{\partial x_i} \right) \qquad i,k=x,y,z
\end{equation}%
is the strain tensor. Inserting $\mbox{\boldmath $u$}_{\mbox{\boldmath $q$}}$, 
Eq.\ (\ref{uq}), into this
formula and imposing the condition $E(\mbox{\boldmath $u$}_{\mbox{\boldmath $q$}}) 
= \hbar \omega_q$
determines the normalization as
\begin{mathletters}\label{norm}
\begin{equation}\label{norm1}
C_q \equiv C = \frac{1}{L} \sqrt{\frac{\hbar}{\rho v_s a}} \, ,
\end{equation}
with a numerical factor
\begin{eqnarray}\label{norm2}
a(\alpha)& = &
f^3-2f+\frac{1}{\kappa_l}-\frac{\alpha^2 \xi^2}{\kappa_l} \\
&& +\frac{1}{\xi^2} \left[
\frac{(1+\kappa_l^2)^2}{2\kappa_l} +\kappa_l(1+f^2)
-2f(1+\kappa_l\kappa_t) 
\right] 
\, . \nonumber
\end{eqnarray}%
\end{mathletters}%
Equations (\ref{norm}) show that the normalization leads merely to a constant
prefactor, i.~e., in contrast to the case of bulk phonons,
$C$ does not introduce a further dependence on the wave number $q$.

We are now in a position to quantize the displacement field 
$\mbox{\boldmath $u$}_{\mbox{\boldmath $q$}}$
[Eqs.\ (\ref{uq})]
of a SAW. According to the familiar rules, we define
the phonon annihilation and creation operators $b_{\mbox{\boldmath $q$}}$ and
$b_{\mbox{\boldmath $q$}}^\dagger$ and
find for the
complete wave field the expression
\begin{equation}\label{uqcomplete}
\mbox{\boldmath $u$}(\mbox{\boldmath $r$},t) =
C \sum_{\mbox{\boldmath $q$}} \left[ b_{\mbox{\boldmath $q$}} 
e^{i(\mbox{\boldmath $qR$} -\omega_q t)}
\mbox{\boldmath $v$}_{\mbox{\boldmath $q$}}(z)
+h.c. \right] .
\end{equation}%

\subsection{Deformation potential interaction}

The deformation potential
is proportional to the change in volume, ${\rm div} \mbox{\boldmath $u$}$,
which an infinitesimal volume element undergoes due to the wave
\cite{Gantmakher87}.
Introducing an electron-phonon interaction constant $\Xi$,
the Hamiltonian of the deformation potential
can be written as
\begin{equation}\label{hdefgen}
H_{DA} = \Xi {\rm div} \mbox{\boldmath $u$}(\mbox{\boldmath $r$},t) \, .
\end{equation}%
The spread of the transversal component of the electron wave function
as well as the distance $d$ of the 2DEG
from the surface are small compared to $q^{-1}$.
Thus, in evaluating Eq.\ (\ref{hdefgen}),
one can set all exponentials in $\mbox{\boldmath $v$}_{\mbox{\boldmath $q$}}(z)$, Eq.\ (\ref{uq2}),
equal to 1.

Conveniently, the electron-phonon interaction Hamiltonian can be written in the
general form
\begin{equation}\label{hdefdet}
H = \frac{1}{L} \sum_{\mbox{\boldmath $q$}} 
\gamma_{\mbox{\boldmath $q$}} e^{i\mbox{\boldmath $q$}\mbox{\boldmath $R$}} b_{\mbox{\boldmath $q$}}
+ h.c. \, .
\end{equation}%
For a deformation potential interaction we derive from Eqs.\ 
(\ref{uqcomplete}) and (\ref{hdefgen})
the interaction vertex
\begin{equation}\label{vertdp}
\gamma_{\mbox{\boldmath $q$}}^{DA} = \sqrt{\frac{\hbar}{\rho v_s a}} \, \alpha \xi^2 \Xi q \, .
\end{equation}%
Following a notation introduced in Ref.\ \onlinecite{Gantmakher87}, the
electron-phonon interaction constant $\Xi$ can be replaced by a
nominal scattering time $\tau_{DA}$. This gives
\begin{equation}\label{tauda}
(\gamma_{\mbox{\boldmath $q$}}^{DA})^2 =
a_{DA}
\frac{\hbar^2 v_s q^2}{p_\circ^3 \tau_{DA}} \, ,
\end{equation}%
where $\hbar p_\circ = (2m^* \hbar \omega_\circ)^{1/2}$
and $a_{DA}=2\pi\alpha \xi^2/a$. $\omega_\circ$
is the frequency of longitudinal
optical phonons and $m^* $ is the effective mass of the electrons.

\subsection{Piezoelectric interaction}

Along with the deformation potential interaction, the piezoelectric electron-phonon
interaction
appears in crystals which lack a center of symmetry, cf.\ for example
Ref.~\onlinecite{Gantmakher87}. In this case, an elastic wave
leads to a polarization \mbox{\boldmath $P$} of the lattice,
\begin{equation}\label{pj}
P_j = \sum\limits_{k,l} \tilde{\beta}_{jkl} u_{kl} \, ,
\end{equation}%
where $\tilde{\beta}_{jkl}$ is the tensor of the piezoelectric moduli.
The corresponding interaction Hamiltonian follows from the electric potential
$\varphi(\mbox{\boldmath $r$}, t)$ associated with the polarization and reads
\begin{equation}\label{hpiezo}
H_{PA} = e \varphi( \mbox{\boldmath $r$},t) \, .
\end{equation}%
The polarization and the electric potential are related to one another
via Poisson's equation
\begin{equation}\label{poisson}
{\rm div } \mbox{\boldmath $D$} = \varepsilon_\circ {\rm div} 
(4 \pi \mbox{\boldmath $P$} -{\rm grad} \varphi) = 0 \, ,
\end{equation}%
where $\mbox{\boldmath $D$}$ is the dielectric displacement and 
$\varepsilon_\circ  \approx 12.8$ is the dielectric constant of GaAs.

In the case of interest here, the general expression (\ref{pj}) is simplified
because the GaAs-samples used in experiments are cubic crystals 
and a crystal cut
is chosen [the (100) surface]
where the surface is spanned by two lattice axes \cite{Wixforth89}.
Then, the tensor $\tilde{\beta}_{jkl}$ has only components in which all three
indices $j$, $k$, $l$ differ from each other and all components are equal to
 $\beta/8\pi$.
Hence, Eq.\ (\ref{pj}) reduces to
\begin{equation}\label{pola}
P_x=(4\pi)^{-1}\beta u_{yz}, \quad P_y=(4\pi)^{-1}\beta u_{zx} ,
\quad P_z=(4\pi)^{-1}\beta u_{xy} \, .
\end{equation}%
Substituting the displacement field (\ref{uqcomplete}) into Eq.\ (\ref{strain})
for the strain tensor yields the polarization (\ref{pola}).
Then the Poisson equation (\ref{poisson}) for $\varphi$ can
be solved most easily by a Fourier transform
in the $x$-$y$-plane, leading to
\begin{equation}\label{poft}
\left[ \frac{\partial^2}{\partial z^2} -q^2 \right] \varphi(z,t) =
\beta C q_x q_y  e^{-i\omega_q t}
\left[
-3\kappa_l e^{-\kappa_l qz} + f(1+2 \kappa_t^2) e^{-\kappa_t qz}
\right] + c.c. \, .
\end{equation}%
The solution of this inhomogeneous differential equation
can be constructed in the usual way.
Discarding the exponentially increasing term $e^{qz}$,
one obtains that every mode with a wave vector \mbox{\boldmath $q$} is associated with an
electric potential
\begin{equation}\label{phiq}
\varphi_{\mbox{\boldmath $q$}}(\mbox{\boldmath $r$},t) = \beta \xi^{-2} C \hat{q}_x \hat{q}_y
e^{i(\mbox{\boldmath $q$}\mbox{\boldmath $R$} - \omega_q t)} \left\{
3\kappa_l \alpha^{-1} e^{-\kappa_l q z} -
f(1+2\kappa_t^2) e^{-\kappa_t q z}
+c_1 e^{-qz}
\right\} + c.c. ,
\end{equation}%
where $\hat{q}_x = \mbox{\boldmath $q$}\hat{\mbox{\boldmath $x$}}/q$
and $\hat{q}_y = \mbox{\boldmath $q$}\hat{\mbox{\boldmath $y$}}/q$.
We note that for the geometry under consideration
the total number of decay lengths for the elastic displacements and the
electric potential is three,
cf.\ Ref.~\onlinecite{Mayer95} for comments on the general case.
$c_1$ is a constant of integration which is determined by the boundary
conditions at the surface $z=0$.
In view of the experiments, we assume that the surface of the crystal is an
`electrically free' boundary \cite{Farnell78} to vacuum.
That is, the normal component of the dielectric displacement, Eq.\ (\ref{poisson}),
and the parallel
components of the electric field are continuous at the surface,
\begin{mathletters}\label{bcond}
\begin{equation}\label{bcond1}
4\pi\varepsilon_\circ P_z -\varepsilon_\circ  \frac{\partial}{\partial z} \varphi|_{z=+0} =
-\frac{\partial}{\partial z} \varphi|_{z=-0} \, ,
\end{equation}%
\begin{equation}\label{bcond2}
\frac{\partial}{\partial \mbox{\boldmath $R$}} \varphi|_{z=+0}
= \frac{\partial}{\partial \mbox{\boldmath $R$}} \varphi|_{z=-0} \, .
\end{equation}%
\end{mathletters}%
Note that the boundary conditions (\ref{bcond}) differ from the
requirement $\varphi_{z=0} =0$
for a sample which is covered with a thin metallic film.
An appropriate ansatz for the electric potential outside of the crystal
($z<0$) is $\varphi=c_2 
e^{i(\mbox{\boldmath $q$}\mbox{\boldmath $R$}- \omega_q t)} e^{qz}$.
Substituting this ansatz and the general solution (\ref{phiq}) for $z>0$
into the boundary conditions (\ref{bcond}) yields that the constant
of integration is 
\begin{equation}\label{coninte}
c_1 = \frac{1}{2\bar\varepsilon}
\left[
-3 \kappa_l \alpha^{-1} (1+ \kappa_l \varepsilon_\circ )
+ f(1+2 \kappa_t^2)(1+\kappa_t \varepsilon_\circ )
- \varepsilon_\circ \xi^2 (1-f\kappa_t)
\right] ,
\end{equation}%
where $\bar\varepsilon=(\varepsilon_\circ +1 )/2$ is the average
 of the dielectric constants of GaAs and the space above the sample
surface (vacuum), respectively.
For large values of $\varepsilon_\circ$, $\varepsilon_\circ \gg 1$,
this result
coincides with the one which follows from the approximate boundary
condition $\frac{\partial}{\partial z} \varphi|_{z=-0} =0$, cf.\
Eq.\ (\ref{bcond1}).
The electric potential (\ref{phiq}) associated with a single displacement mode
is now completely determined. 

Assigning the amplitudes $b_{\mbox{\boldmath $q$}}$ and
$b_{\mbox{\boldmath $q$}}^\dagger$
to the first and second term in Eq.\ (\ref{phiq}), respectively,
summing over all wave vectors, and
introducing the result
into the Hamiltonian (\ref{hpiezo}), the piezoelectric 
vertex  [see Eq.\ (\ref{hdefdet})] becomes
\begin{equation}\label{vertpi}
\gamma_{\mbox{\boldmath $q$}}^{PA} =
\sqrt{\frac{\hbar}{\rho v_s a}} \,
\beta e \xi^{-2} \frac{\varepsilon_\circ}{2\bar\varepsilon}
\hat{q}_x\hat{q}_y
\left[
3 \kappa_l \alpha^{-1} (1-\kappa_l)
-f(1+2 \kappa_t^2)(1-\kappa_t)
-\xi^2 (1-f\kappa_t)
\right] ,
\end{equation}%
where we set $z=0$ in $\varphi_{\mbox{\boldmath $q$}}(\mbox{\boldmath $r$},t)$, Eq.\ (\ref{phiq}).
Obviously, the strongest piezoelectric interaction occurs when the
SAW is propagated along a diagonal direction ($\hat{q}_x\hat{q}_y =
\pm \frac{1}{2}$). In the experiments just this piezoelectric
active direction [$\mbox{\boldmath $q$} \| [011]$] is chosen.
In terms of a nominal time $\tau_{PA}$ [cf.\ Eq.\ (\ref{tauda})] the
interaction vertex reads
\begin{equation}\label{taupa}
(\gamma_{\mbox{\boldmath $q$}}^{PA})^2 =
a_{PA} (\hat{q}_x\hat{q}_y)^2 \frac{ \hbar^2 v_s}{p_\circ \tau_{PA}} \, ,
\end{equation}%
where all the numerical quantities are absorbed in the
prefactor $a_{PA}$.

\subsection{Discussion of the interaction vertices}
Let us compare the results  
for the interaction vertices $\gamma_{\mbox{\boldmath $q$}}$ in the Hamiltonian
(\ref{hdefdet})
with those valid for 3D bulk phonons (or a fictitious 2D phonon system).
There are two significant differences.
First, the interaction vertices for
SAW's have a different
dependence on the wave vector: $|\gamma_{SAW}|^2$ exhibits an additional
factor $q$ compared to $|\gamma_{bulk}|^2$. This applies to both
the deformation potential and the piezoelectric interaction.
Consequently, the use of the SAW interaction vertices in calculations
of various physical quantities can give rise to results which deviate from
those which are based on the assumption of two- or three-dimensional
phonon systems.
Second, for surface phonons, the deformation potential and 
the piezoelectric interaction are {\it in} phase.
This is in contrast to the case of bulk phonons where these
vertices are {\it out} of phase, i.~e.\ they
contribute additively to $|\gamma_{bulk}|^2=
|\gamma_{bulk}^{DA} + \gamma_{bulk}^{PA}|^2 =
|\gamma_{bulk}^{DA}|^2 + |\gamma_{bulk}^{PA}|^2$,
see for example Ref.\ \onlinecite{Gantmakher87}.
The absolute value squared of $\gamma_{\mbox{\boldmath $q$}}$ 
is the relevant quantity which determines
the total electron-phonon interaction.
Clearly, the `out of phase' or the `in phase' property is of importance only
when the interaction vertices for the deformation potential and 
the piezoelectric interaction
are of the same order of magnitude.
This depends on the wavelength of the SAW because $\gamma^{PA}$,
Eq.\ (\ref{vertpi}), does not depend on the 
magnitude of $q$ whereas $\gamma^{DA}$, Eq.\
(\ref{vertdp}), increases linearly with $q$.
In the case of GaAs, the relative strength of these two interaction mechanisms
is thus
$\gamma^{DA}/\gamma^{PA} \approx q 10^{-7} {\rm cm} $
where we have used the numerical values given below.
Thus, for the range of wavelengths used in recent experiments
on the attenuation of a SAW in GaAs-samples (see, for example,
Refs.~\onlinecite{Rampton92,Wixforth86,Wixforth89,Schenstrom88,Willet90} 
and \onlinecite{Guillion91}),
the deformation potential scattering can be neglected in comparison with
the piezoelectric interaction, except for propagation 
along $\left\langle 100 \right\rangle$ directions. 
This result corroborates very well with the fact that
the experimental findings
could be explained in terms of the piezoelectric electron-phonon
coupling alone \cite{Wixforth89}.

For easy reference, we summarize the numerical values for the parameters
appearing in the interaction vertices $\gamma_{\mbox{\boldmath $q$}}$. 
For GaAs, $c_l\approx 5\times 10^5$cm/s,
$c_t \approx 3\times 10^5$cm/s and, hence, $\alpha=0.36$. 
The corresponding solution of the algebraic equation \cite{Landau70}
for $\xi$ is $\xi\approx 0.9$. Substituting these values in Eq.\ (\ref{norm})
yields $a=1.4$. Using $\tau_{DA}=4$ps, $\tau_{PA}=8$ps,
$\hbar \omega_\circ=421$K (this corresponds to
$\Xi=7.4 $eV, $e\beta=2.4 \times 10^7$eV/cm),
and $\rho = 5.3$g/cm${}^3$
(see Ref.\ \onlinecite{Gantmakher87}),
we obtain $\gamma_{\mbox{\boldmath $q$}}^{DA} =5.6\times 10^{-17}q\,$eVcm${}^2$ and
$\gamma_{\mbox{\boldmath $q$}}^{PA} = 3.7 \hat{q}_x\hat{q}_y 10^{-10}$eV cm.

The above calculations are restricted, with respect to the piezoelectric
interaction, to cubic crystals and a particular crystal cut.
It is straightforward, however, to perform calculations for different
crystals or surfaces along the same lines.

\section{Electron states and percolation theory}

Consider a 2DEG in a strong magnetic field  $B$ perpendicular to
the plane of the 2DEG
and in a smooth potential $V(\mbox{\boldmath $R$})$ (see, for instance, the paper
by Trugman\cite{Trugman83} and references therein).
The potential $V(\mbox{\boldmath $R$})$ is assumed to vary slowly on
the scale of the magnetic length  $l_B=\sqrt{c\hbar/ eB}$.
Electron-electron interactions are neglected.
The wave function $\Psi(\mbox{\boldmath $R$})$ of a state with energy
$\epsilon$ is appreciable only in the vicinity of
an equipotential line (EL) of the potential $V(\mbox{\boldmath $R$})=\epsilon$.
The width of the wave functions perpendicular to the EL is $l_B$.
Explicitly, the electron states of the $n$-th Landau level (LL)
can be approximated, in the limit $B \rightarrow \infty$, by
\begin{equation}\label{Psi}
\Psi(\mbox{\boldmath $R$}) \equiv \Psi(u,v) = 
[{\cal T}v_D(u,v)]^{-1/2} \chi_n(v) e^{i\varphi(u,v)} \, .
\end{equation}%
(We have omitted the part $\Psi(z)$ of the wave function
which corresponds to the lowest occupied subband perpendicular
to the plane of the 2DEG.)
The orthogonal variables $u$ and $v$ parametrize the distances
along and perpendicular to the EL, respectively.
The function $\chi_n(v)$ is the $n$th harmonic oscillator function.
Below, we shall restrict ourselves to the lowest Landau level,
i.~e.\ $n=0$. In Eq.\ (\ref{Psi}),
$\varphi(u,v)$ is a gauge-dependent phase, and
${\cal T}$ is the period associated with one revolution
around the EL of an electron moving with the drift
velocity $v_D$.
That is,
\begin{equation}\label{periodt}
{\cal T} =
\oint \! du \, \frac{1}{v_D(u,v)} \, ,
\qquad v_D = |\nabla V|l_B^2/\hbar \, .
\end{equation}%
For the wave function (\ref{Psi}) to be single-valued,
$\varphi(u,v)$ has to change by an integral multiple of $2\pi$
around an EL. This condition leads to the quantization 
of the allowed constant-energy lines. In other words, only
a discrete sequence of EL's corresponds to the electron eigenstates.
Two adjacent eigenstates enclose an area $2\pi l_B^2$
and are, on the average, a distance $\Delta v \simeq l_B^2/{\cal L}$ apart,
where ${\cal L}$ is the length of one of the EL's.
An important quantity is the difference
$\hbar \omega_{{\cal T}}$ of the corresponding
eigenenergies, where the frequency $\omega_{{\cal T}}$ is determined by
\begin{equation}\label{omtdef}
\omega_{{\cal T}} = 2 \pi / {\cal T} \, .
\end{equation}%
The quasi-classical description of the electron states that
we have outlined above is a valid approximation when\cite{Trugman83}
\begin{equation}\label{cond}
l_B/r_c \ll 1 \quad  \mbox{\rm and } \quad
m^* |\nabla V(\mbox{\boldmath $R$})| l_B^3/ \hbar^2 \ll 1 \, ,
\end{equation}%
where $r_c$ is the local radius of curvature of the EL
and $m^*$ is the effective electron mass.
The first condition is related to the smoothness of the potential, 
while the second one prevents the mixing of different LL's.
Additionally, one should keep in mind that quantum tunneling\cite{Fertig87} between classical
EL's is important when $|\epsilon|$ is smaller than
$\Delta (l_B/\Lambda)^2$.

In what follows $V(\mbox{\boldmath $R$})$ is a smooth {\it random} potential.
The potential is assumed to be Gaussian,
with
\begin{equation}\label{correlator}
\left\langle V(\mbox{\boldmath $R$}) V(0) \right \rangle =\Delta^2\phi(R/\Lambda) \, ,
\end{equation}%
where $\Delta=\sqrt{\left\langle V^2 \right \rangle}$ defines its amplitude
and $\Lambda$ its correlation length. ($\Delta$ determines also
the broadening of the LL.)
The zero of energy is chosen such that 
$\left\langle V(\mbox{\boldmath $R$}) \right \rangle =0$, i.~e.,
the energy $\epsilon$ is measured from the center of
the lowest LL. 
Using $\Delta$ and $\Lambda$, we can rewrite the conditions
(\ref{cond}) in the form
\begin{equation}\label{condr}
l_B/\Lambda \ll 1 \quad  \mbox{\rm and } \quad
\frac{\Delta}{\hbar \omega_c} \, \frac{l_B}{\Lambda} \ll 1 \, ,
\end{equation}%
where $\omega_c = eB/cm^*$ is the cyclotron frequency.
These conditions are fulfilled, for example, for the
following experimental values: $B=10$T ($l_B=8$nm,
$\omega_c= 2\pi\times 4.2$THz), $\Delta=1$meV and $\Lambda= 50$nm.
The separation between two consecutive LL's is much larger
than their broadening, $\hbar\omega_c/\Delta \approx 60$.

As discussed in Ref.\ \onlinecite{Iordansky96},
most EL's with $\epsilon$ in the tail of the LL, i.~e.
$|\epsilon|\gg \Delta$, have diameters ${\cal D}$
which are small compared to $\Lambda$ and their length ${\cal L}$
is of the order of ${\cal D}$.
When the energy
approaches the center of the LL, the size of the  EL's grows
\cite{Mehr88}.
In particular, for  $|\epsilon|\simeq \Delta$,
${\cal L}\simeq{\cal D}\simeq\Lambda$ holds for most of the EL's.
Such EL's will be denoted as `standard'.
A further reduction of $|\epsilon|$ does not lead 
to an increase in the size of almost all EL's, i.~e.\ most of them remain
standard ones. However, a minority of the EL's merges and forms
large extended EL's with diameters
${\cal D}\gg\Lambda$.
The structure of these extended EL's is described by
percolation theory
\cite{Stauffer92} when the energies $|\epsilon|$ are near
the percolation threshold $\epsilon=0$ 
(or $|\epsilon|/\Delta \ll 1$).
The subsequent calculations show 
that just this range of energies is of interest justifying the use
of the percolation picture.

An extended EL has a fractal structure 
which is reflected in the relation between its length and diameter
\cite{Stauffer92}
\begin{eqnarray}
{\cal L}\simeq \Lambda ({\cal D}/\Lambda)^{2/\alpha} \, ,
\label{ldrela}
\end{eqnarray}
where $\alpha=8/7$ is the scaling exponent.
(For the definition of
the perimeter of discrete percolation clusters and the transition
to continuum percolation
see, e.~g., Refs.\ \onlinecite{Grossman86} and \onlinecite{Mehr88}.)
An extended EL can be viewed
as a self-avoiding random walk path with steps of length $\Lambda$.
Indeed, $2/\alpha =7/4$ is close to the value 2 which
applies to a simple random walk. The exponent is less than 2
due to the self-avoiding nature of the EL.

The distribution of extended EL's of a given energy $\epsilon$
is described by one scale, the so-called
critical diameter
\begin{eqnarray}
{\cal D}_{c}(\epsilon)\simeq \Lambda ({|\epsilon|/ \Delta})^{-\nu} \, ,
\label{dcrit}
\end{eqnarray}
which is considered to be the localization length
in the semiclassical theory. The
scaling index is $\nu=4/3$.
EL's with diameters ${\cal D}\gg {\cal D}_{c}(\epsilon) $ are exponentially
rare, while the probability to find an extended EL with
a diameter $\Lambda \ll {\cal D}\ll {\cal D}_{c}(\epsilon) $ is proportional to
${\cal D}^{-\rho}$, where $\rho=3$.

An electron that moves on an extended EL (${\cal L} \gg \Lambda$)
experiences different
regions of the random potential. During one revolution,
the drift velocity $v_D(u,v)$ [see Eq.\ (\ref{periodt})] 
follows the varying slope of the potential $V(\mbox{\boldmath $R$})$ and
takes on many different values.
In other words, the motion on an extended EL corresponds to
an averaging process with respect to $v_D(u,v)$.
It is therefore reasonable to introduce
an average drift velocity\cite{Iordansky96} $\bar v_D$, 
defined by ${\cal T}={\cal L}/\bar v_D$,
that is assumed to be independent of
the length of the EL under consideration.
The dependence on the energy $\epsilon$ can be generally excluded
since $V(\mbox{\boldmath $R$})$ and $\nabla V(\mbox{\boldmath $R$}) 
\sim v_D(\mbox{\boldmath $R$})$
are statistically independent for a Gaussian distribution
\cite{Longuet57}.
Consequently, the energy level spacing (\ref{omtdef}) associated with
the extended EL's is a function of ${\cal L}$ alone and
Eq.\ (\ref{omtdef})
can be written conveniently in the form
\begin{equation}\label{omegat}
\hbar \omega_{\cal T} ({\cal L }) = 
\hbar \Omega \frac{\Lambda }{{\cal L }}  \, ,
\end{equation}%
where
\begin{equation}\label{Omega}
\hbar\Omega=
\hbar \frac{2\pi \bar v_D}{\Lambda}
\simeq
\frac{\Delta l_B^2}{\Lambda^2} \, .
\end{equation}%
The frequency $\Omega$
gives by order of magnitude the level spacing
for standard EL's since it is associated with 
the revolution around an EL with
${\cal L} \simeq \Lambda$. The corresponding frequencies for
the extended EL's are lower.
The lowest frequencies belong to the
longest EL's which have the critical length ${\cal L}_c$
corresponding to the critical diameter ${\cal D}_c$ (\ref{dcrit}).
From Eqs.\ (\ref{ldrela}) and (\ref{dcrit})
\begin{equation}\label{areac}
{\cal L }_c (\epsilon) \simeq \Lambda |\Delta/ \epsilon|^{2 \nu/\alpha} \, .
\end{equation}%

Below, we shall use the
distribution of the EL's with respect to their lengths.
Let $L^2f_{\epsilon}({\cal L})d{\cal L}$ be the number of EL's
with energy $\epsilon$ and a length between ${\cal L}$ and
${\cal L}+d{\cal L}$.
The normalization of this distribution can be found by
equating the total average length of the EL's in
an area of size $L^2$
to the result given in the literature
(see Sec.\ III.A in Ref.\ \onlinecite{Longuet57} or Ref.\
\onlinecite{Isichenko92})
\begin{equation}\label{normfe}
L^2 \int_{0}^{\infty}d{\cal L}{\cal L}f_{\epsilon}({\cal L})=
\frac{L^2}{2 \Lambda}
[- \phi''(0)]^{1/2}
\exp[-{\epsilon^2\over 2\Delta^2}] \, ,
\end{equation}
where $\phi$ is defined in Eq.\ (\ref{correlator}).

While the distribution of the standard EL's is not
really known, percolation theory gives the following ansatz
\cite{Stauffer92}
for the distribution of extended EL's
\begin{equation}\label{feext}
f_{\epsilon}({\cal L})d{\cal L} = C_\epsilon 
\left( \frac{{\cal L}}{\Lambda} \right)^{-[{\alpha\over 2}(\rho-1)+1]}
G\left({{\cal L}\over{\cal L}_{c}(\epsilon)}\right)d{\cal L} \, ,
\qquad {\cal L}\gg \Lambda \, ,
\end{equation}
where $C_\epsilon$ is the normalization constant and
$G(\zeta)$ is a function which is exponentially small
for $\zeta\gg 1$ and of order unity for $\zeta\ll 1$.
Hence, $G$ yields a smooth cut-off of the distribution
for ${\cal L} > {\cal L}_c$, where ${\cal L}_c$ 
is defined in Eq.\ (\ref{areac}).
An additional, energy-independent cut-off appears in a finite sample.
Here, the size $L$ of the system restricts the critical 
diameter ${\cal D}_c$ (\ref{dcrit})
to values such that ${\cal D}_c \lesssim L$. This translates into 
${\cal L} \lesssim \Lambda (L/\Lambda)^{2/\alpha}$,
using Eq.\ (\ref{ldrela}).
Thus, in a finite system, the critical length ${\cal L}_c(\epsilon)$
in Eq.\ (\ref{feext}) should be replaced by min$\{ {\cal L}_c(\epsilon),
\Lambda (L/\Lambda)^{2/\alpha}\}$.

To find the normalization constant $C_\epsilon$ let us decompose
the  normalization integral (\ref{normfe})
into $\int_0^\Lambda + \int_\Lambda^\infty$ and estimate both terms
of this decomposition. The second integral can be estimated
from the distribution (\ref{feext}) of extended EL's.
With the value ${\alpha\over 2}(\rho-1)+1=15/7$,
this integral is determined by its lower limit $\Lambda$
and is of the order $(L \Lambda)^2$.
Using a reasonable
ansatz for the distribution of standard and short EL's (for example
$f_{\epsilon}({\cal L})=const.$),
the first integral is determined by its upper limit $\Lambda$
and is again of the order  $(L \Lambda)^2$. Thus, 
the total normalization integral is also of this order.
Up to a factor of order unity, the normalization constant
$C_\epsilon$ is then given by $\Lambda^{-3}$.
The numerical factor can be absorbed in $G$ leading to
the following distribution function for extended EL's
\begin{equation}\label{fel}
f_{\epsilon}({\cal L})d{\cal L}=
{1\over\Lambda^3} \left({{\cal L}\over \Lambda}\right)
^{-[{\alpha\over 2}(\rho-1)+1]}
G\left({{\cal L}\over{\cal L}_{c}(\epsilon)}\right)d{\cal L} \, .
\end{equation}%
We note that the above estimates confirm
that the majority of EL's belongs to the standard ones with
${\cal L} \simeq \Lambda$,
since these EL's are relevant in the normalization integral.

\section{Matrix elements}
Emission and absorption of phonons are associated with 
electronic transitions 
with energy transfer $\hbar \omega_q$.
We have seen in the previous section that
the separation in energy between two consecutive EL's is
given by $\hbar \omega_{\cal T}$ [Eq.\ (\ref{omtdef})].
Thus, {\it real} transitions are generally restricted to EL's
for which $\omega_{\cal T} \le \omega_q$. 
For the parameters used above for conditions (\ref{condr}),
the frequency $\Omega$, Eq.\ (\ref{Omega}),
is about $2\pi \times 10$GHz, whereas the frequencies
of the SAW's used in experiments vary typically in the range
$\omega_q=2\pi\times 1$MHz $\div$ 1GHz.
We therefore conclude that only extended EL's
for which $\omega_{\cal T} = \Omega \Lambda/{\cal L} \ll \Omega$
[see Eq.\ (\ref{omegat})] contribute to the sound absorption.
Thus, the matrix elements of the interaction Hamiltonian (\ref{hdefdet})
\begin{equation}\label{mael}
{\cal M}_{if}^{\pm \mbox{\boldmath $q$}}=
\frac{1}{L} \gamma_{\mbox{\boldmath $q$}} M_{if}^{\pm \mbox{\boldmath $q$}} 
\equiv\frac{1}{L} \gamma_{\mbox{\boldmath $q$}}
\left\langle f | e^{\pm i \mbox{\boldmath $q$} \mbox{\boldmath $R$}} | i \right\rangle \, ,
\end{equation}%
where $|\left. \! i\right\rangle$ and
$|\left. \! f\right\rangle$ denote the initial and final wave functions
of the form (\ref{Psi}), have to be calculated for extended trajectories.
This calculation has been performed in Ref.\ \onlinecite{Iordansky96}.
The matrix element, averaged over all trajectories with the same period ${\cal T}$
and the same energy $\epsilon$,
reads
\begin{equation}\label{me1}
\left\langle |M_{if}^{\pm\mbox{\boldmath $q$}}|^2 \right\rangle_{\epsilon, {\cal T}} = 
c q^2 \Lambda^2 (\hbar\Omega)^{\alpha}
\frac{\hbar\omega_{\cal T}}{|\epsilon_f - \epsilon_i|^{\alpha+1}}
\quad \mbox{\rm for} \quad |\epsilon_f - \epsilon_i| \lesssim \hbar\Omega \, ,
\end{equation}%
where $c$ is a numerical factor of order unity.
The matrix element is valid under the assumptions
$q\Lambda \ll v_D/v_s \ll (q\Lambda)^{-3/4}$,
where the exponent $3/4$ follows from $(2-\alpha)/\alpha$ with $\alpha=8/7$,
cf.\ Eq.\ (\ref{ldrela}). Clearly, these inequalities imply
$q\Lambda \ll 1$.

For $|\epsilon_f - \epsilon_i| \gg \hbar\Omega$, the matrix element
$\left\langle |M_{if}^{\pm\mbox{\boldmath $q$}}|^2 \right\rangle_{\epsilon, {\cal T}}$
is exponentially small.
This implies that transitions occur only within the
lowest LL and that transitions to other LL's can be neglected
($\hbar\omega_c \gg \Delta \gg \hbar\Omega$).
It is also assumed that the initial and final states are close
to one another in real
space: In order that $(\chi_0)_i$ and $(\chi_0)_f$
will overlap, the separation in real space, $\Delta v$,
should satisfy $\Delta v \lesssim l_B$.
The condition $|\epsilon_f - \epsilon_i| \lesssim \hbar\Omega$ is even
more restrictive. This can be seen in the following way.
As mentioned above, the mean distance in real space
between two adjacent EL's is given by $l_B^2/{\cal L}$.
Hence, the distance between the two states $i$ and $f$
is of order
$(l_B^2/{\cal L}) |\epsilon_f - \epsilon_i|/\hbar \omega_{\cal T}$.
The maximum of this expression is found for the largest
allowed energy difference $|\epsilon_f - \epsilon_i| \simeq \hbar \Omega$.
Using the definition of $\omega_{\cal T}$ in Eq.\ (\ref{omegat})
and the estimate for $\Omega$ in Eq.\ (\ref{Omega}),
the corresponding maximum distance in real space is 
found to be $l_B^2/\Lambda \ll l_B$,
cf.\ the inequalities (\ref{condr}).

While the sound absorption is due to transitions between extended
states, the calculation of the dielectric function
(see the next section) necessitates also matrix elements
between standard EL's.
(Transitions between a standard EL and an extended EL are 
exponentially rare
due to their large separation in space.)
Since the majority of the EL's belongs to the standard ones,
one might even expect that the standard EL's 
dominate the dielectric function. This is not the case, 
as is shown below.

Noting that for typical phonon
wave vectors (e.~g.\ $q\approx 10^4$cm${}^{-1}$),
one has $q\Lambda \ll 1 $,
the matrix element for standard EL's with ${\cal L} \simeq \Lambda$
can be approximated by
\begin{equation}\label{meshort1}
\left\langle f | e^{i \mbox{\boldmath $q$} \mbox{\boldmath $R$}} | i \right\rangle
= e^{i \mbox{\boldmath $q$} \mbox{\boldmath $R$}_i}
\left\langle f | e^{i \mbox{\boldmath $q$} 
(\mbox{\boldmath $R$}-\mbox{\boldmath $R$}_i)} | i \right\rangle
\approx
e^{i \mbox{\boldmath $q$} \mbox{\boldmath $R$}_i} i\mbox{\boldmath $q$}
\left\langle f | \mbox{\boldmath $R$}-\mbox{\boldmath $R$}_i | i \right\rangle \, ,
\end{equation}%
where the zero-order term in the expansion of the exponential function
disappears due to the orthogonality of the two states $i$ and $f$.
The vector $\mbox{\boldmath $R$}_i$ denotes some point in the vicinity of
the $i$th EL.
The matrix element on the right-hand-side
of Eq.\ (\ref{meshort1}) is of order $\Lambda$.
Hence, 
for transitions between two standard EL's
\begin{equation}\label{meshort2}
\left\langle |M_{if}^{\pm\mbox{\boldmath $q$}}|^2 \right\rangle_{\epsilon, {\cal T}}
\approx q^2\Lambda^2 \, ,
\end{equation}%
where it is understood that the EL's $i$ and $f$ are very 
close in real space and in energy; otherwise the matrix element is exponentially
small. The first condition guarantees the overlap of  the 
wave functions $\chi_0^{i(f)}$, see Eq.\ (\ref{Psi}).
The second one is necessary to ensure that the integrand of the $u$-integration along
the perimeter of the EL's is not a fast oscillating function.
Result (\ref{meshort2}) agrees essentially with Eq.\ (\ref{me1}) replacing there
the energy difference $|\epsilon_f-\epsilon_i|$ and the level
spacing $\hbar \omega_{\cal T}$ by the value $\hbar \Omega$ appropriate
for standard EL's.

\section{The dielectric function}

The matrix element (\ref{mael})
includes the screening
of the electron-phonon interaction due to the lattice 
[Eq.\ (\ref{vertpi})]. 
The screening arising from the 2DEG can be accounted for by
renormalizing the matrix element
\begin{equation}\label{renor}
|{\cal M}_{if}^{\pm \mbox{\boldmath $q$}}|^2 \rightarrow
\frac{ |{\cal M}_{if}^{\pm \mbox{\boldmath $q$}}|^2 }{|\varepsilon(\omega_q, q)|^2} \, ,
\end{equation}%
where $\varepsilon(\omega,q)$ is the dielectric function
of the 2DEG.
For a nearly half-filled LL, 
the dielectric function 
can be calculated assuming linear screening\cite{Wulf88}. That is,
the change in the electron density resulting from a small applied
potential is proportional to the strength of the perturbing potential.
Indeed, one can estimate that for the SAW intensities used in experiments
the electron density oscillates only weakly around its average
value, see for example Ref.\ \onlinecite{Esslinger94}.
The assumption of linear screening leads to the general expression
\begin{equation}\label{dfdef}
\varepsilon(\omega, q)=
1 + \frac{2 \pi e^2}{\bar\varepsilon q} \Pi(\omega, q) \, ,
\end{equation}%
where
\begin{equation}\label{dencor}
\Pi(\omega, q)=
\frac{1}{L^2}
\sum\limits_{i \neq f}
\frac{f(\epsilon_i) -f(\epsilon_f)}{\epsilon_f -\epsilon_i -\hbar \omega -i0}
|M_{if}^{\mbox{\boldmath $q$}}|^2
\, ,
\end{equation}%
and $\bar\varepsilon$  is defined
in Eq.\ (\ref{coninte}).

To evaluate $\Pi$ explicitly, we
transform the sum $\sum_{i\neq f}$ in Eq.\ (\ref{dencor})
into $\sum_{i < f}$, where $i<f$ means $\epsilon_i <\epsilon_f$.
Let us first focus on the case of zero temperature, i.~e.\
all levels below the Fermi energy $\epsilon_F$
are occupied, $f(\epsilon_i)=1$,
whereas all levels above $\epsilon_F$ are empty, $f(\epsilon_f)=0$.
Then
\begin{equation}\label{pitzero}
\Pi(\omega, q)=\frac{2}{ L^2}
\sum\limits_{i<f}
\frac{\epsilon_f-\epsilon_i}
{(\epsilon_f -\epsilon_i)^2 -(\hbar \omega +i0)^2}
|M_{if}^{\mbox{\boldmath $q$}} |^2
\, .
\end{equation}%
In order to yield an appreciable matrix element,
the EL's corresponding to $\epsilon_i$ and $\epsilon_f$ must
be close (in real space and in energy) to an EL
with $\epsilon=\epsilon_{F}$. This `Fermi' EL (FEL) need not to be
an electron state.
We can therefore represent the summation over states
in Eq.\ (\ref{pitzero})
as a sum over EL's near a certain FEL and then sum over all FEL's.
In the first sum the states are distributed nearly
equidistantly with an energy spacing
$\hbar\omega_{{\cal T}}\approx const.$, cf.\ below.
In the summation over FEL's we may first sum 
over the FEL's with the same
period ${\cal T}$.
Since these EL's are situated in different regions of the
random potential,
this summation is equivalent to an average
of the matrix element over FEL's with the same period. Thus,
the averaged matrix elements (\ref{me1}) and (\ref{meshort2}) 
for extended and standard EL's, respectively,
can be substituted in Eq.\ (\ref{pitzero}).

We begin with the contribution of the extended EL's to $\Pi$.
As we shall see, this is the dominant contribution.
It is easy to see that the energy spacing for
the relevant states near a fixed FEL is given 
by the value of $\hbar \omega_{\cal T}$
at the Fermi energy. To this end, we have to calculate the change
in $\omega_{\cal T}$ arising from a change in the energy of the EL by at most
$\hbar\Omega$ [see Eq.\ (\ref{me1})]. 
Since the frequency $\omega_{\cal T}$ for the extended
EL's is merely a function of ${\cal L}$, one has
$\Delta \omega_{\cal T}/\omega_{\cal T} \simeq
\Delta {\cal L}/{\cal L} \simeq \Delta {\cal A}/{\cal A}$,
where ${\cal A}$ is the area enclosed by the EL.
To get the second equality, we have used 
${\cal L} \simeq \Lambda ({\cal A}/\Lambda^2 )^\lambda$,
with\cite{Stauffer92} $\lambda=12/13$. The change in the enclosed
area is given by $2\pi l_B^2 \Omega/\omega_{\cal T}$ and thus
$\Delta \omega_{\cal T}/\omega_{\cal T} \simeq
(l_B^2/\Lambda^2) (\Lambda/{\cal L})^{1/\lambda-1} \ll 1$.
Consequently, the sum over EL's  which are near a given FEL
can be simplified by introducing an
explicit representation for the energies
\begin{equation}\label{energyeps}
\epsilon_f - \epsilon_i = (m-n) \hbar\omega_{\cal T} \, .
\end{equation}%
The integers $m$ and $n$ are subject to the restrictions
$|m-n| \lesssim \Omega/\omega_{\cal T} $ [see Eq.\ (\ref{me1})] and
$m-n \neq 0$.
Using the representation (\ref{energyeps}) and the matrix
element (\ref{me1}), the double sum over states near one FEL
in Eq.\ (\ref{pitzero})
can be reduced to a sum over $s=m-n$ and one obtains
\begin{equation}\label{pzero1}
\sum_{i<f}
\frac {1}{(\epsilon_f-\epsilon_i)^2-(\hbar\omega+i0)^2}
\frac {\hbar\omega_{{\cal T}}}{|\epsilon_f-\epsilon_i|^{\alpha}}=
\left( \frac{x}{\hbar\omega} \right)^{\alpha+1} S(x)
\end{equation}%
where $x=\omega/\omega_{{\cal T}}$ and
\begin{equation}\label{sums}
S(x)\equiv
\sum_{s=1}^{\infty}
{1\over s^2-(x+i0)^2}{1\over s^{\alpha-1}} \, .
\end{equation}
We have replaced the upper limit in the sum by infinity, since the relevant
$s$ are of order $x\ll \Omega/\omega_{{\cal T}} $ and the
above mentioned restriction for $|m-n|$ can be neglected.
In other words, the EL's which contribute significantly are separated
in energy by $\hbar\omega$.

For extended states, using Eq.\ (\ref{omegat}),
$x=(\omega/\Omega)({\cal L}/\Lambda) $,
and hence the contribution to $\Pi(\omega, q)$
from a FEL is a function of its length alone. As a result the total
$\Pi(\omega, q)$ can be written as a sum over all lengths.
Using the distribution function $f_\epsilon({\cal L})$ 
given in Eq.\ (\ref{fel}),
we find
\begin{eqnarray}\label{pzero2}
\Pi(\omega, q; \epsilon_F)& = & 2c \,
\frac{(q\Lambda)^2 \Omega^\alpha}{\hbar \omega^{\alpha+1}}
\int d{\cal L} \, f_{\epsilon_F}({\cal L})
x^{\alpha+1}S(x)
\nonumber\\
& = &
2c \, \frac{q^2}{\hbar \omega} \,  H(y_F) \, ,
\end{eqnarray}
where
\begin{equation}\label{yfdef}
y_F= \frac{\Omega \Lambda}{\omega {\cal L}_c(\epsilon_F)}
=
\left| \frac{\epsilon_F}{\epsilon_\omega} \right|^{2\nu/\alpha}
\qquad {\rm and} \qquad
\epsilon_\omega=
\Delta \left( \frac{\omega}{\Omega} \right)^{\alpha/2\nu}
\, .
\end{equation}%
In a finite system of size $L$, $y_F$ has to be replaced by the maximum
of $y_F$ and $y_L\equiv(\Omega/\omega)(\Lambda/L)^{2/\alpha}$,
see the discussion following Eq.\ (\ref{feext}).
Using the explicit form for the distribution function $f$ with $\rho=3$
yields
\begin{equation}\label{fh}
H(y)=\int_{0}^{\infty}dxG(xy)S(x) \, .
\end{equation}
The quantity $\epsilon_\omega$ 
has a very intuitive interpretation: it is the energy
at which the energy level spacing 
$\hbar\omega_{\cal T}({\cal L}_c(\epsilon))$ for an EL with the critical length
is equal to the phonon energy $\hbar\omega$ of the SAW.
In other words, $\epsilon_\omega$ determines the absorption threshold
in the sense that real transitions occur only for
$|\epsilon_F| \lesssim \epsilon_\omega$. This is reflected in the 
imaginary part of $\Pi$, calculated below.

The real and imaginary parts of the function $H$ are given by 
\begin{mathletters}\label{hf}
\begin{equation}\label{hfreal}
{\rm Re} H(y)=\int_0^{\infty}dx \, G(xy)
\sum_{s=1}^{\infty}\frac{1}{s^{\alpha-1}}
\frac{P}{s^2-x^2} \, ,
\end{equation}%
\begin{equation}\label{hfimag}
{\rm Im} H(y)=\frac{\pi}{2}\sum_{s=1}^{\infty}\frac{1}{s^{\alpha}}
G(sy) \, ,
\end{equation}%
\end{mathletters}%
where $P$ denotes the principal part of the integral.
The behavior of $H(y)$ in an infinite 
system is shown in Fig.\ 1.
The $y$-axis in Fig.\ 1
has been scaled in terms of $y^{3/7}$ corresponding to the
dependence of $H$ on the Fermi energy, see
Eq.\ (\ref{yfdef}).
The limiting behaviors for large and small arguments 
are given by
\begin{equation}\label{limith}
\begin{array}{rclrclrcl}
{\rm Re} H & \approx & \zeta(1+\alpha)/y =1.5/y, &
\qquad {\rm Im}H & \approx & (\pi/2)G(y), & \qquad y& \gg &1\, , \\
{\rm Re} H & \simeq & y^{\alpha-1}, & 
\qquad {\rm Im}H & \simeq & 1-y^{\alpha-1}, &
\qquad y &\ll &1\, , \\
{\rm Re} H & = & 0, & \qquad {\rm Im}H & = & 
(\pi/2)\zeta(\alpha)=11.9, & \qquad y& =&0 \, ,
\end{array}
\end{equation}
where $\zeta(x)=\sum_{s=1}^\infty s^{-x}$. 
In order to discuss the analytic expressions of $H(y)$
we note that the sum in Eq.\ (\ref{hfimag})
and the integral in Eq.\ (\ref{hfreal})
are truncated at $s$ or $x$ of order $1/y$,
as implied by $G$, Eq.\ (\ref{feext}).
The imaginary part (\ref{hfimag})
is therefore exponentially small
$\sim G(y)$ for
$y\gg 1$ and of order unity in the opposite
case. Thus, the sum over $s$ increases with
decreasing $y$ and approaches
its maximum as $y$ goes to zero.
In a finite system, $y$ is restricted from below by $y_L$ 
imposing an upper limit $1/y_L$ on the sum. This leads to a
smaller maximum value of Im$H$.  
As for the real part of $H(y)$, we are able to study
the limiting cases.
For both $y \gg 1$ and $y \ll 1$,
Re$H(y)$ goes to zero according to a power law.
In the intermediate region $y \lesssim 1$ but not $y \ll 1$,
Re$H$ is slowly varying and of order unity. 
In a finite system, the real part approaches a small but finite value
as $y$ goes to $y_L$.

Up to this point, the evaluation of $\Pi$, Eq.\ (\ref{dencor}),
has been performed for zero temperature, $T=0$.
The calculation of $\Pi$ for finite temperatures such
that $T \gg \hbar\omega$ can be done
along the same lines. Therefore, we give only a brief
description.
The calculations for $T=0$ have shown that the
real and the imaginary parts
of $H$ reach a value of order unity when the Fermi energy becomes
of order $\epsilon_\omega$.
This energy range corresponds to the contribution of EL's 
with a length
${\cal L}_\omega \equiv {\cal L}_c(\epsilon_\omega) \simeq \Lambda (\Omega/\omega)$
and an energy level spacing of order $\hbar\omega$
to the transition processes.
Therefore these extended EL's
give the dominant contributions to $\Pi$, Eq.\ (\ref{dencor}).
This remains true for finite
temperatures. However, now the states $i$ and $f$ need not be in
the immediate vicinity of
the FEL's. Instead, we can fix some initial state $i$ and consider
transitions to final states above, $\epsilon_f > \epsilon_i$, and
below, $\epsilon_f < \epsilon_i$, the chosen one
(essentially in a range $T$ around $\epsilon_F$). To do this,
we use the representation (\ref{energyeps}) where $\omega_{\cal T}$
now refers to the energy $\epsilon_i$.
Expanding the Fermi distribution $f(\epsilon_f)$ in Eq.\
(\ref{dencor}) around $\epsilon_i$ leads
to $-\partial f(\epsilon_i)/\partial \epsilon_i$
(for the relevant transitions $|\epsilon_f-\epsilon_i|$ is much
smaller than $T$).
The sum over the energies $\epsilon_f$ is reduced to twice
the expression (\ref{sums}). Hence
\begin{equation}\label{tnz2}
\Pi(\omega,q)=
2c\left( \frac{q\Lambda}{L} \right)^2 (\hbar \Omega)^\alpha
\sum\limits_i \left(- \frac{\partial f}{\partial \epsilon_i} \right)
(\hbar \omega_{\cal T})^{-\alpha} 
S\left(\frac{\omega+i0}{\omega_{\cal T}}\right) \, .
\end{equation}%
The sum over the initial states $i$ comprises
a summation over all EL's with the same energy $\epsilon_i$ but
with different lengths ${\cal L}$, and a summation over
$\epsilon_i$. The first sum can be again replaced by an integral
using the distribution function $f_\epsilon({\cal L})$, Eq.\ (\ref{fel}).
Then, the order of summation and integration is inverted
to obtain
\begin{equation}\label{tnonzero}
\sum\limits_{\epsilon_i} \int_0^\infty d{\cal L}
\quad
\rightarrow 
\quad
\int_0^\infty d{\cal L} \sum\limits_{|\epsilon_i| \lesssim \epsilon_c({\cal L})}
\quad
\rightarrow 
\quad
\int_0^\infty d{\cal L} \int_{-\infty}^{\infty}
\frac{d\epsilon_i}{\hbar\omega_{\cal T}({\cal L})} \, .
\end{equation}%
The condition $|\epsilon_i| \lesssim \epsilon_c({\cal L})$ ensures
that only those initial states which possess
a critical length ${\cal L}_c$ equal to or larger than ${\cal L}_c(\epsilon_i)$
are included.
Here $\epsilon_c({\cal L}) \simeq \Delta (\Lambda/{\cal L})^{\alpha/2\nu}$,
see Eq.\ (\ref{areac}).
However, the dominant contributions to $\sum_{\epsilon_i}$ follow
from a particular group of EL's, rendering this condition unnecessary.
Taking also into account that the relevant EL's have an energy level spacing 
which is small compared to the thermal
energy $\hbar\omega/T \ll 1$,
and that the number of states per energy interval is given by
$(\hbar\omega_{\cal T})^{-1}$,
the sum over $\epsilon_i$ can be replaced by the integral given on the
right-hand-side of Eq.\ (\ref{tnonzero}).
The limits of integration have
been extended with negligible error.
The resulting integral over ${\cal L}$ coincides with the right-hand-side
of Eq.\ (\ref{pzero2}) except for the value of the energy:
instead of the fixed Fermi energy $\epsilon_F$
there appears now the variable $\epsilon_i$.
Thus, we finally arrive at
\begin{equation}\label{pzerot}
\Pi(\omega,q)= 
\int_{-\infty}^{\infty} d\epsilon \,
\left(- \frac{\partial f}{\partial \epsilon} \right) \Pi(\omega,q; \epsilon) \, ,
\end{equation}%
where we have dropped the index $i$ of $\epsilon_i$.
This equation shows that a finite temperature leads to
an average of the $T=0$ result 
over energies within an interval of order $T$
around the Fermi level.
Since the function $\Pi(\omega,q; \epsilon)$ varies
on the scale $\epsilon_\omega$, finite temperature effects are negligible
if $T \ll \epsilon_\omega$. This is the condition for
Eq.\ (\ref{pzero2}) to hold. For $T \gtrsim \epsilon_\omega$,
the width of $\Pi(\omega,q)$ as function
of the Fermi energy increases with temperature,
i.~e.\ the behavior of $\Pi$ deviates substantially from the
$T=0$ result.
In the following we assume $T < \epsilon_\omega$.

Substituting the $T=0$ result for $\Pi(\omega,q)$ [Eq.\ (\ref{pzero2}), which is
based on transitions between extended EL's]
in Eq.\ (\ref{dfdef})
yields for the dielectric function
\begin{equation}\label{dfgen}
\varepsilon(\omega,q)=
1+\frac{2\pi e^2}{\bar\varepsilon}\frac{q}{\hbar\omega}
2cH(y_F)
\end{equation}
and, for the renormalization of the matrix element in Eq.\
(\ref{renor}),
\begin{equation}\label{dfspec}
\varepsilon(\omega_q,q)=
1+\frac{e^2}{\bar\varepsilon \hbar v_s}\, 4\pi c H(y_F) \, .
\end{equation}
The contribution to the dielectric function resulting from
standard EL's is derived below, see Eq.\ (\ref{chishort}). The comparison
of that result with Eqs.\ (\ref{dfgen}) and (\ref{dfspec})
shows that the dielectric function is essentially given by
the contribution due to extended EL's,
whereas the influence of transitions between standard EL's
can be neglected.
We consider therefore Eq.\ (\ref{dfspec}) as the final result
for $\varepsilon(\omega_q,q)$.

The dielectric function (\ref{dfspec}) renormalizes the matrix element
(\ref{renor}) via the expression
$|\varepsilon(\omega_q, q)|^2$. The dependence of $|\varepsilon|^2$ on the
Fermi energy is given by $|H|^2$, Eqs.\ (\ref{hf}).
The latter is of order unity for $|\epsilon_F| \lesssim \epsilon_\omega$
and decreases for larger values of the Fermi energy
according to the power law $(\epsilon_\omega/|\epsilon_F|)^{4\nu/\alpha}$,
$4\nu/\alpha=14/3$.
Thus, the magnitude of $\varepsilon$, Eq.\ (\ref{dfspec}), is determined by the
large dimensionless parameter
$e^2/\bar\varepsilon v_s \hbar$ ($\approx 110$ for GaAs). 
This is the ratio
of the electrostatic energy of two electrons a distance $q^{-1}$
apart and the energy of a surface phonon, 
$(e^2 q/\bar\varepsilon)(\hbar \omega_q)^{-1}$.

Let us now show that the contribution of the standard EL's to
the dielectric function is negligible.
We start afresh from the expression (\ref{pitzero})
for $\Pi$, substituting the matrix element 
(\ref{meshort2}) for transitions between
standard EL's.
The energy difference between two standard
EL's is of order $\hbar\Omega$, Eq.\ (\ref{omegat}), i.~e.\ much
larger than $\hbar\omega$. The latter can thus
be neglected in comparison with $\epsilon_f -\epsilon_i$ in Eq.\
(\ref{pitzero}). Then the imaginary term $i0$ can be dropped, 
as no real transitions can occur.
The sum over all final states $f$ leads merely to a factor
of order one, since only a few EL's in the immediate vicinity
of the initial EL contribute to the matrix element (\ref{meshort2}).
The remaining sum over the initial states counts the 
standard EL's which are just below the Fermi level. 
The number of these EL's is essentially the number of all FEL's,
because the number of very short (${\cal L} \ll \Lambda$)
and very long (${\cal L} \gg \Lambda$) EL's is negligibly small
for Fermi energies near the center of the LL.
The required quantity may therefore be deduced from Eq.\ (\ref{normfe})
which states that the total length of all EL's is, up to a numerical
factor, equal to $L^2/\Lambda=\Lambda (L^2/\Lambda^2)$. Since the mean length
of all EL's is of order $\Lambda$, the number of FEL's in a system
of size $L$ is of order $L^2/\Lambda^2$.

Collecting these results, we obtain 
the contribution due to standard EL's
\begin{equation}\label{chishort}
\Pi(\omega, q) \simeq
\frac{q^2}{\hbar \Omega} \qquad {\rm and} \qquad \varepsilon(\omega, q) -1 \simeq
\frac{e^2 q}{\bar\varepsilon \hbar\Omega} \, .
\end{equation}%
It can be shown that this estimate is valid 
independent of the ratio $\hbar\Omega/T$
as long as max$\{ \hbar \Omega, T \} \ll \Delta$.
The comparison of Eq.\ (\ref{chishort})
with Eq.\ (\ref{dfgen}) shows that the contribution 
of the standard EL's to the dielectric function is $\omega/\Omega$ times
smaller than the term resulting from the extended EL's.

It is instructive to consider briefly an alternative derivation of the
dielectric function which reproduces correctly the order of magnitude
$|\varepsilon| \simeq e^2/\bar\varepsilon v_s \hbar$.
This derivation relies on the fact that the motion of an electron on a fractal
trajectory can be considered as a self-avoiding random walk
with single steps of length $\Lambda$.
In fact, the relation between the diameter and the length of an
extended EL is similar to what one would expect for
a simple random walk, cf.\ Eq.\ (\ref{ldrela}).
For the diffusive regime, the density correlator $\Pi$, Eq.\ (\ref{dencor}),
is given by
\begin{equation}\label{dencorr}
\Pi(\omega, q) =
g_F \frac{Dq^2}{-i\omega +Dq^2} \, ,
\end{equation}%
where $D$ is the diffusion constant and $g_F$ the density of states
at the Fermi level.
In our case, we can assume $D \simeq v_D \Lambda$.
The density of states
of the LL is given by
$g_F \simeq (\Delta l_B^2)^{-1}$ for $\epsilon_F \ll \Delta$.
Substituting these quantities into Eq.\ (\ref{dfdef})
yields
\begin{equation}\label{rpafinal}
\varepsilon(\omega_q, q) - 1
\simeq i \frac{e^2}{\bar\varepsilon v_s \hbar} \, ,
\end{equation}%
where the term $Dq^2$ has been neglected compared to $-i\omega_q$
in the denominator of the density correlator.
Interestingly, this approach predicts an essentially imaginary
result for $\varepsilon -1$ which agrees with
the behavior of Eq.\ (\ref{dfspec}) for $|\epsilon_F| \rightarrow 0$,
i.~e.\ in the case when some the EL's become arbitrarily long.

\section{Surface acoustic wave attenuation}

The intensity of the SAW decreases due to absorption of phonons by the 2DEG
with the distance $x$ as exp$(-\Gamma x)$.
The attenuation coefficient $\Gamma$ can be expressed in terms of the sound
velocity
$v_s$ [cf.\ Eq.\ (\ref{omega})] and the life time $\tau(\mbox{\boldmath $q$})$ as
$\Gamma= (v_s \tau(\mbox{\boldmath $q$}))^{-1}$, where $\tau(\mbox{\boldmath $q$})$
is defined by the rate equation
\begin{equation}\label{rate1}
\dot{N}_{\mbox{\boldmath $q$}} = - 
\frac{1}{\tau(\mbox{\boldmath $q$})} N_{\mbox{\boldmath $q$}} \, .
\end{equation}%
Here $N_{\mbox{\boldmath $q$}}$ is the phonon occupation number.
The net change in $N_{\mbox{\boldmath $q$}}$ is given by
\begin{eqnarray}
\label{rate2}
\dot{N}_{\mbox{\boldmath $q$}} & =& 
\frac{2 \pi}{\hbar |\varepsilon(\omega_q,q)|^2}
\,\, \sum_{i \neq f}
f(\epsilon_i)(1-f(\epsilon_f))  \nonumber \\
&& \times [
|{\cal M}_{if}^{-\mbox{\boldmath $q$}}|^2 (N_{\mbox{\boldmath $q$}} +1) 
\delta(\epsilon_i - \epsilon_f -\hbar \omega_q )
- |{\cal M}_{if}^{+\mbox{\boldmath $q$}}|^2 N_{\mbox{\boldmath $q$}} 
\delta(\epsilon_i - \epsilon_f +\hbar \omega_q )] ,
\end{eqnarray}%
where $f(\epsilon)$ is the Fermi distribution function and
${\cal M}_{if}^{\mp \mbox{\boldmath $q$}}$
are the unscreened matrix elements (\ref{mael}) for 
emission or absorption of a phonon with
wave vector $\mbox{\boldmath $q$}$. 
For a SAW induced by interdigital transducers, the phonon occupation number
$N_{\mbox{\boldmath $q$}}$ is macroscopically large. 
The difference between $N_{\mbox{\boldmath $q$}}+1$ and 
$N_{\mbox{\boldmath $q$}}$ is therefore
negligible. Combining Eqs.\ (\ref{rate1}) and (\ref{rate2}) yields
\begin{equation}\label{rate3}
\frac{1}{\tau(\mbox{\boldmath $q$})}=
\frac{2 \pi}{\hbar |\varepsilon(\omega_q,q)|^2}
\sum_{i \neq f} 
|{\cal M}_{if}^{\mbox{\boldmath $q$}}|^2 [f(\epsilon_i)-f(\epsilon_f)]
\delta(\epsilon_i - \epsilon_f +\hbar \omega_q ) \, .
\end{equation}%
Replacing the $\delta$-function by the imaginary part
of $-\pi^{-1} [\epsilon_i - \epsilon_f +\hbar \omega_q +i0]^{-1}$,
we find
\begin{equation}\label{rate4}
\frac{1}{\tau(\mbox{\boldmath $q$})}=
\frac{2}{\hbar}
\frac{|\gamma_{{\mbox{\boldmath $q$}}}|^2}{|\varepsilon(\omega_q,q)|^2}
{\rm Im} \Pi(\omega_q,q)_{\omega_q>0} \, ,
\end{equation}%
where $\Pi$ is defined by Eq.\ (\ref{dencor}).

Using the zero-temperature results for $\Pi$
and the dielectric function, 
Eqs.\ (\ref{pzero2}) and (\ref{dfspec}),
respectively, as well as the relation 
between the life time $\tau(\mbox{\boldmath $q$})$ and the attenuation coefficient,
we find
\begin{equation}\label{gscr}
\Gamma =
\Gamma_q \Phi(y_F)
\, , \qquad 
y_F=|\epsilon_F/\epsilon_\omega|^{2\nu/\alpha} \, ,
\end{equation}%
with
\begin{equation}\label{gq}
\Gamma_q=\frac{1}{4\pi^2c {\rm Im}H(0)} 
|\gamma_{\mbox{\boldmath $q$}}|^2 \frac{q \bar\varepsilon^2}{e^4}
\, , \qquad
{\rm and} \qquad
\Phi(y)= {\rm Im}H(0) \frac{{\rm Im}H(y)}{|H(y)|^2} \, ,
\end{equation}%
where the 
term 1 in the expression (\ref{dfspec}) for $\varepsilon$ has 
been neglected. The function $\Phi(y)$ is defined such that $\Phi(0)=1$
[since Re$H(0)=0$, cf.\ Eqs.\ (\ref{limith})],
i.~e., $\Gamma_q$ coincides with the attenuation coefficient
at the center of the LL, $\Gamma(\epsilon_F=0)=\Gamma_q$.
We begin with the discussion of the magnitude of $\Gamma_q$ and consider
the function $\Phi(y)$ afterwards.

Substituting the expressions (\ref{tauda}) and (\ref{taupa})
for the interaction vertices
in Eqs.\ (\ref{gq}) yields
\begin{mathletters}\label{gmag}
\begin{equation}\label{gmagda}
(\Gamma_q)_{DA} = \frac{a_{DA} }{4\pi^2c {\rm Im}H(0)}
\frac{q^3}{v_s p_\circ^3 \tau_{DA}}
\left( \frac{\bar\varepsilon v_s \hbar}{e^2} \right)^2
= 2.6\times 10^{-21} q^3 \mbox{\rm cm}^2 \, ,
\end{equation}%
\begin{equation}\label{gmagpa}
(\Gamma_q)_{PA} = \frac{a_{PA} }{4\pi^2c {\rm Im}H(0)}
\frac{q}{v_s p_\circ \tau_{PA}}
\left( \frac{\bar\varepsilon v_s \hbar}{e^2} \right)^2
= 8.0\times 10^{-6} q \, . \qquad \,\, {}
\end{equation}%
\end{mathletters}%
That is, despite 
the fractal structure of the extended EL's on which these results
are based, the frequency dependence of the magnitude of $\Gamma$
is simple and is not characterized by scaling exponents.
Moreover, $\Gamma_q$ is independent of the magnetic field and
the parameters $\Lambda$ and $\Delta$ of the random potential.
The numerical values on the right-hand-side of Eqs.\ (\ref{gmag})
have been calculated replacing the parameters $p_\circ, \tau_{DA},$ etc.\
by theirs values given in Sec.\ II and assuming $c=1$. 
For a finite system, Im$H(0)$ [see Eqs.\ (\ref{limith})]
has to be replaced by Im$H(y_L)<{\rm Im}H(0)$,
see the discussion following Eq.\ (\ref{yfdef}),
leading to an increase of the attenuation coefficient at the center
of the LL.

The function $\Phi(y_F)$ in Eq.\ (\ref{gscr}) accounts for the dependence
of $\Gamma$ on the Fermi energy (or the filling factor $\bar\nu$ or
the magnetic field $B$). This dependence is determined
by the ratio of $|\epsilon_F|$ and
the energy $\epsilon_\omega=\Delta (\omega_q/\Omega)^{\alpha/2\nu}$
as follows.
The absorption of the SAW is
very small when the Fermi energy is far from the center
of the LL $|\epsilon_F| \gg \epsilon_\omega$, i.~e.\ $y_F\gg 1$.
A strong increase of $\Phi$ and, hence, of $\Gamma$ occurs when $|\epsilon_F|$
is reduced to $|\epsilon_F| \approx \epsilon_\omega$. 
In this region the number of occupied extended EL's
with an energy level spacing 
$\hbar \omega_{\cal T}({\cal L}) \lesssim \hbar \omega_q$,
Eq.\ (\ref{omegat}), undergoes the change from an exponentially small quantity to
some power-law function of ${\cal L}^{-1}$.
(Nevertheless, the number of these states is negligible compared
to the majority of EL's with ${\cal L} \simeq \Lambda$.)
A further rise of the absorption is prevented by the enhanced
screening $\sim ({\rm Im}H)^2$ at $|\epsilon_F|\ll \epsilon_\omega$
which even reduces $\Gamma$ as the Fermi energy goes to zero.
This results in a shallow double-peak structure
with a cusp at the center of the LL.
In fact, if we use the limiting forms of 
$H(y\ll 1)$, Eqs.\ (\ref{limith}), we find 
$d\Gamma/d\epsilon_F \simeq {\rm sgn}(\epsilon_F)/|\epsilon_F|^{2\nu(2-\alpha)/\alpha}$.
We believe therefore that the double-peak structure of $\Gamma(\epsilon_F)$
is independent of the function $G(z)$ used to describe the 
exponential cut-off of the extended EL's, see the discussion following
Eq.\ (\ref{feext}).
The maxima of $\Gamma(\epsilon_F)$ are located
near $\pm \epsilon_\omega$, see Fig.\ 2.
It is clear, however, that such a particular feature as the cusp has to be considered
with caution, for it is exclusively based on the quasiclassical model for
the electron states. Quantum tunneling between
critical trajectories may modify this result. It is worth to note that
the tunneling 'band' of width \cite{Fertig87} $\Omega$ [Eq.\ (\ref{Omega})]
around $\epsilon=0$
[cf.\ the discussion after Eq.\ (\ref{cond})]
is narrow compared to the characteristic energy range $\epsilon_\omega$;
indeed, $\Omega/\epsilon_\omega \approx (l_B/\Lambda)\sqrt{\omega/\Delta} \ll 1$.
One may speculate that the absorption coefficient 
is only weakly affected by quantum tunneling. Indeed, most of the EL's
contributing to $\Gamma$ cannot be connected by low saddle points with
transmission coefficients of order unity, and so quantum tunneling in-between
them is insignificant.

To simplify the estimates, we rewrite $\epsilon_\omega$
in the form
\begin{equation}\label{epoest}
\epsilon_\omega= 0.3 {\rm meV}
\left( \frac{ \Delta}{{\rm 1meV}} \right)^{4/7}
\left( \frac{\omega_q}{2\pi\times 1{\rm GHz}} \right)^{3/7}
\left( \frac{\Lambda}{50{\rm nm}} \right)^{6/7}
\left( \frac{ B}{5{\rm T}} \right)^{3/7} \, .
\end{equation}%
As discussed in Sec.\ V, the zero-temperature result (\ref{pzero2}) for
$\Pi$ remains valid for finite temperatures such that
$T \ll \epsilon_\omega$.
This is also the condition for Eq.\ (\ref{gscr}) to hold.
Using the values for $\Delta$ and $\Lambda$ given above and
$\omega_q=2\pi\times 100$MHz, we obtain $\epsilon_\omega \approx 1$K.
For temperatures of the order of or larger
than $\epsilon_\omega$, the attenuation coefficient is found
from Eq.\ (\ref{rate4}) using expression (\ref{pzerot}) in
the calculation of the dielectric function and Im$\Pi$.
Two results for $\Gamma$ at finite temperatures are shown in Fig.\
2. With increasing temperature the minimum of the attenuation
near the center of the LL is reduced and the
absorption peak becomes broader. The increasing
magnitude of $\Gamma$ results from the significant broadening
of the imaginary part of $\Pi$ and the reduced screening,
see Eq.\ (\ref{rate4}).
For $T \gtrsim \epsilon_\omega$, the magnitude of $\Gamma$ and the width of
the absorption region are strongly temperature dependent.

The dependence of $\Gamma$ on the SAW frequency $\omega_q$
is shown in Fig.\ 3.
The curve is calculated for the low temperature regime
$T \ll \epsilon_\omega$ and the piezoelectric electron-phonon
interaction. The attenuation coefficient has been written in the
form $\Gamma=\Gamma_F (\omega_q/\omega_F) \Phi(\omega_F/\omega_q)$,
with $\Gamma_F=(\Gamma_q/q)(\omega_F/v_s)$, 
$\omega_F=\Omega |\epsilon_F/\Delta|^{2\nu/\alpha}$.
(Note that $\Gamma_q/q$ does not depend on frequency.)
The Fermi energy is fixed to some
value $\epsilon_F \ll \Delta$
and defines the smallest level spacing $\hbar\omega_F$ for
extended EL's. Consequently,
$\omega_q \simeq \omega_F$ marks the onset of
strong SAW attenuation. For high frequencies $\omega_q \gg \omega_F$,
the attenuation coefficient increases linearly with frequency.
This is just the behavior predicted by the classical description
of sound absorption for piezoelectric interaction,
see Eq.\ (\ref{gsig}) below.

For $T\ll\epsilon_\omega$, 
the width of the absorption region 
is determined by
$|\epsilon_F|\simeq \epsilon_\omega$. This is merely the condition
for real transitions to occur and is neither associated with
the interaction vertices $\gamma_{\mbox{\boldmath $q$}}$ nor is a consequence
of the matrix element (\ref{me1}) whose derivation
is based on the particular assumption $\bar{\nu} \approx 1/2$.
We believe therefore that this result applies to other
half-integer filling factors $\bar{\nu}$ as well.
To express the relation $|\epsilon_F|\simeq \epsilon_\omega$
in terms of the filling factor
$\bar\nu=2\pi l_B^2 n$, we write the electron density $n$
as an integral over the Gaussian density of states,
\begin{equation}\label{gauss}
g(\epsilon)=(2\pi)^{-3/2} (l_B^2\Delta)^{-1}
\exp(-\epsilon^2/2\Delta^2) \, ,
\end{equation}%
and the Fermi distribution function $f$,
\begin{equation}\label{nuefint}
\bar \nu(\epsilon_F)=
\frac{1}{\sqrt{2\pi} \Delta}
\int_{-\infty}^\infty d\epsilon \, 
e^{-\epsilon^2/2\Delta^2} f(\epsilon - \epsilon_F) \, ,
\end{equation}%
and expand around the center of the LL with respect to
$|\epsilon_F|/\Delta \ll 1$.
This gives for $T\ll \Delta$
\begin{equation}\label{nuef}
\Delta \bar\nu(\epsilon_F) =
\bar\nu(\epsilon_F) - \bar\nu(0)
= \frac{\epsilon_F}{\sqrt{2\pi} \Delta} \, .
\end{equation}%
Then, the width of the absorption
region is obtained as
\begin{equation}\label{dnu}
|\Delta \bar{\nu}| \simeq
\left( \frac{\omega_q}{\Omega} \right)^{\alpha/2\nu}
\simeq
\left(q \Lambda \frac{v_s}{\bar{v}_D}\right)^{\alpha/2\nu} \, .
\end{equation}%
The exponent is given by $\alpha/2 \nu=\sigma/\lambda=3/7\approx0.42$.
This value agrees with the exponent $\kappa$ which determines the shrinking
of the peaks in the longitudinal conductivity
\cite{Wei92,Aleiner94} $\sigma_{xx}$
as the temperature $T$ goes to zero,
$|\Delta \bar{\nu}| \sim T^\kappa$.
In our case the broadening of the 
absorption peak
arises from the frequency $\omega_q$.
In this sense, $\hbar\omega_q$ may be considered 
as an effective temperature which replaces the real
temperature $T$. Frequency scaling
in the integer quantum Hall regime has been observed
in microwave experiments
\cite{Engel93}. For spin-split LL's, the width of 
the peaks in Re$\sigma_{xx}$ corresponding to different
LL's was found to scale as 
$|\Delta \bar{\nu}| \sim\omega^\kappa$, with
$\kappa\approx 0.41$.

Due to the drift velocity $\bar{v}_D$, the width  (\ref{dnu}) depends
weakly on the absolute value of the magnetic field.
In terms of the filling factor,
Eq.\ (\ref{dnu}) can be written in the form
$|\Delta \bar{\nu}| \approx (n\Lambda^2\hbar \omega_q/\Delta)^{\alpha/2\nu}
(\bar{\nu})^{-\alpha/2\nu}$. Thus, $|\Delta \bar{\nu}|$
is smaller for higher (half-integer) filling factors $\bar{\nu}$.
The width of the absorption region scales with
the phonon wave vector as $|\Delta \bar{\nu}| \sim q^{\alpha/2\nu}$.
In contrast, in the fractional quantum Hall regime,
the width increases linearly with $q$ for $\bar{\nu}=1/2$.
This linear dependence is derived within the composite
Fermion model\cite{Halperin93} and is well confirmed experimentally
\cite{Willet90,Willet94}.

The absorption of SAW's in the integer
quantum Hall regime has also been studied in Ref.\ \onlinecite{Aleiner94}.
These authors determine first the ac-conductivity
of the 2DEG which is then related to the attenuation coefficient
using the equation
\begin{equation}\label{gsig}
\Gamma =
\frac{1}{2} K_{eff}^2 \frac{q\sigma'}{(1+\sigma'')^2+(\sigma')^2} \, ,
\end{equation}%
where
$K_{eff}^2$ represents the effective piezoelectric coupling
constant ($=6.4\times 10^{-4}$ for GaAs\cite{Wixforth89}) and
$\sigma'={\rm Re} \sigma_{xx}(\omega_q,q)/\sigma_M$, 
$\sigma''={\rm Im} \sigma_{xx}(\omega_q,q)/\sigma_M$
and $\sigma_M=v_s\bar\varepsilon/2\pi$.
[Note that Eq.\ (\ref{gsig}) can be obtained from Eq.\
(\ref{gscr}) writing the dielectric function of the 2DEG
in the form $\varepsilon(\omega,q)=1+i\sigma_{xx}(\omega_q,q)/\sigma_M$.]
Assuming that $|\sigma_{xx}| \ll \sigma_M$
for sufficiently high frequencies\cite{Aleiner94},
Eq.\ (\ref{gsig}) reduces to 
$\Gamma(\omega_q,q) \sim q {\rm Re} \sigma_{xx}(\omega_q,q)$.
That is, in this case, the sound absorption
and the longitudinal conductivity are related such that
the width and the shape of their peaks as 
function of $\bar{\nu}$ are identical.
The calculation of the ac-conductivity in Ref.\
\onlinecite{Aleiner94} is based on the concept of
variable-range
hopping between pairs of localized states. For $\hbar\omega_q \gg T$,
the absorption of SAW's is due to resonant phononless transitions of the
electrons from one site of a pair to the other. This mechanism is strongly
affected by the electron-electron interaction \cite{Aleiner94}.
The width of the absorption peak
at half-integer filling factors was found to be
\begin{equation}\label{nuas}
|\Delta \bar{\nu}| \simeq (q\xi_\circ)^{1/\gamma} \, ,
\end{equation}%
where $\gamma\approx 2.3$ is the scaling exponent of the localization length
\begin{equation}\label{xizero}
\xi \simeq \xi_\circ |\bar\nu - \bar\nu(0)|^{-\gamma} \, ,
\end{equation}%
and $\xi_\circ$ is assumed to be of the order of the magnetic length.
[Note the differences between the last equation and the semiclassical
definition of the localization length in Eq.\ (\ref{dcrit}).]
The result of Ref.\ \onlinecite{Aleiner94},
Eq.\ (\ref{nuas}), agrees with our result,
Eq.\ (\ref{dnu}), in both the 
numerical value of the exponent
and the dependence on $q$. However, the width
$|\Delta \bar{\nu}|$ in Eq.\ (\ref{nuas}) exhibits a different 
dependence on the magnetic field, namely
$|\Delta \bar{\nu}| \sim B^{-1/2\gamma}$ in contrast to
$|\Delta \bar{\nu}| \sim B^{\alpha/2\nu}$ predicted by Eq.\
(\ref{dnu}).
The authors of Ref.\ \onlinecite{Aleiner94}
did not give a definite description of the shape of the
absorption peak but rather suggested two scenarios which eventually
lead to a flat peak with a broad maximum or a double-peak,
respectively. Our results support the latter one,
see Fig.\ 2.

\section{Summary}
We have calculated the dielectric function $\varepsilon(\omega,q)$
and the attenuation coefficient $\Gamma$ of a surface acoustic wave
for a 2DEG in a smooth
random potential (with amplitude $\Delta$ and correlation
length $\Lambda$) and a strong magnetic field corresponding to
a filling factor $\bar\nu$ close to $1/2$.
Both quantities become independent of temperature as the
temperature is reduced below a frequency-dependent value
$\epsilon_\omega= \Delta (\omega/\Omega)^{\alpha/2\nu}$,
where $\alpha/2\nu=3/7$, $\Omega=2\pi \bar v_D/\Lambda$ and 
$\bar v_D$ is the average
drift velocity of the electrons
on the equipotential lines of the random potential.
In this low temperature, high frequency regime
(e.~g.\ $\epsilon_\omega \simeq 1$K for
$\omega=2\pi \times 100$MHz), Im$\varepsilon(\omega,q)$ 
and $\Gamma$ are only appreciable
when $\epsilon_F$ is 
within a narrow region around the center of the Landau level, and
Re$\varepsilon(\omega,q)$ decreases according to a power law with
increasing distance from the center.
In particular, the attenuation of the SAW is exponentially small except
for a region whose width 
$|\Delta\bar\nu| \sim \omega^{\alpha/2\nu}$.
This scaling is non-universal because $|\Delta\bar\nu|$
depends on the absolute value of the magnetic field, see
Eq.\ (\ref{dnu}).
The dependence of $\Gamma$ on the Fermi energy
(or the filling factor) yields a double-peak which is centered
at the filling factor $\bar\nu=1/2$, cf.\ Fig.\ 2.
The minimum of the absorption at $\bar\nu=1/2$ results from
the enhanced screening due to the 2DEG, i.~e., 
from the large magnitude of the dielectric function
$|\varepsilon(\omega_q,q)| \simeq e^2/\bar\varepsilon v_s \hbar$,
where $\bar\varepsilon$ is the average of the dielectric constants
of GaAs and vacuum, and $v_s$ is the sound velocity.
The double-peak in $\Gamma$ is most pronounced
for an infinite system
where the critical diameter ${\cal D}_c$, 
Eq.\ (\ref{dcrit}), of the equipotential
lines of the random potential is allowed to take on arbitrarily large values.
A real system of size $L$ restricts the diameter to ${\cal D}_c \lesssim L$
resulting in an increase of the attenuation coefficient 
near the center of the Landau level. 
While this effect is weak for a macroscopic
sample size, a similar but more pronounced effect may arise
from a non-uniform electron density associated with
a spatially varying filling factor.

In the high temperature, low frequency regime, the dielectric
function decreases with rising temperature leading to an increase
of the magnitude of the attenuation coefficient and
a significant increase of the width of the absorption
region around $\bar\nu=1/2$.

\section*{Acknowledgement}

Financial support by the German-Israeli Foundation is gratefully acknowledged.
We thank J.\ Hajdu and D.\ Polyakov for valuable discussions and comments.
One of us (A.~K.) thanks the Deutsche Forschungsgemeinschaft for financial
support and B.~Zingermann for a discussion of some properties of Gaussian
distributions.

\newcommand{\noopsort}[1]{} \newcommand{\printfirst}[2]{#1}
  \newcommand{\singleletter}[1]{#1} \newcommand{\switchargs}[2]{#2#1}

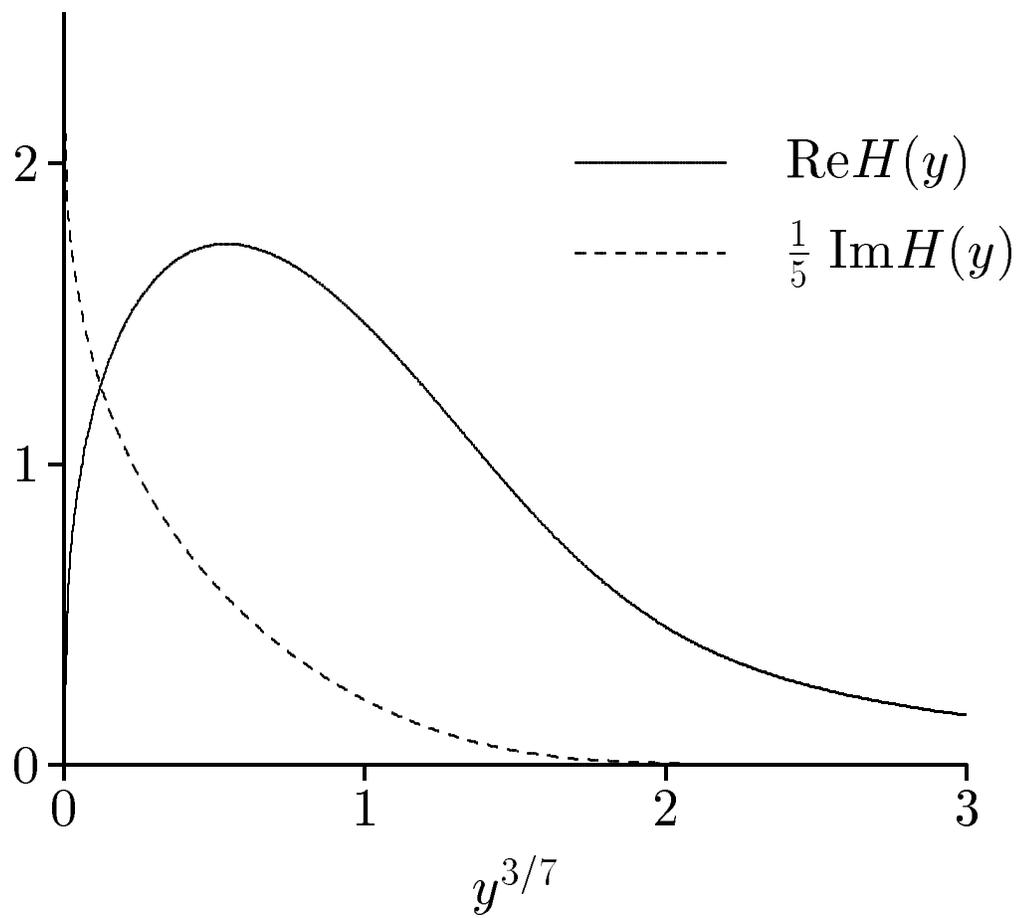
\begin{figure}
\caption{
The real and the imaginary parts of the function $H(y)$ defined in Eqs.\ 
(\protect\ref{hf}). $G(z)$ was replaced by $2/[\exp{(z})+1]$.
The particular choice of the function $G$ has no 
influence on the qualitative behavior of $H$ when
the cut-off introduced by $G$ is smooth enough to wash out all discrete
features of the sums in Eqs.\ (\protect\ref{hf}).
The following figures are based on the curves of $H$ given here.
}
\end{figure}
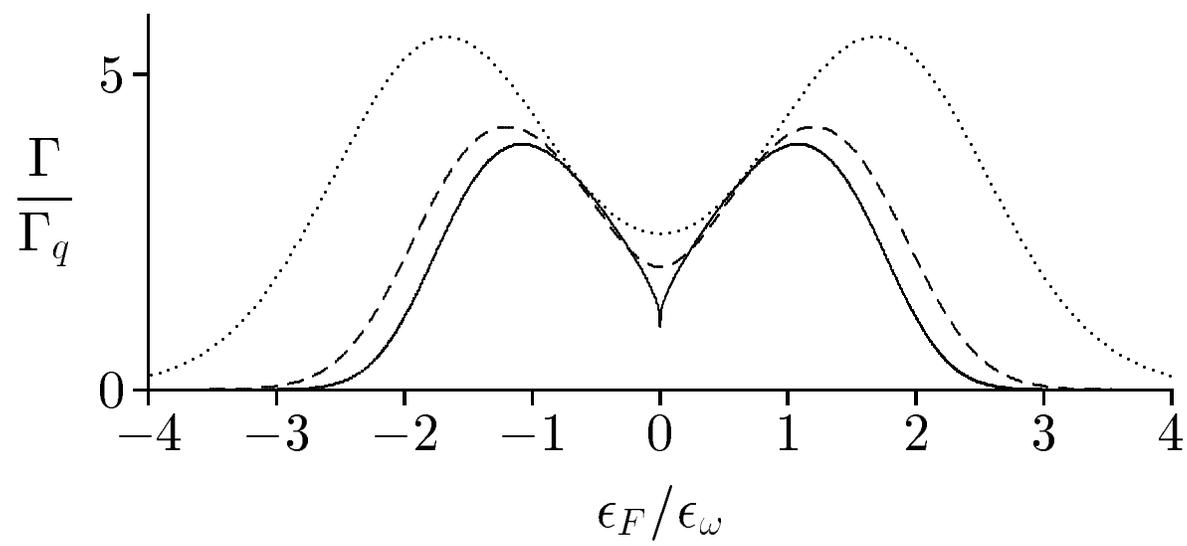
\begin{figure}
\caption{
The attenuation coefficient $\Gamma$, Eq.\ (\protect\ref{gscr}), as a function of
the Fermi energy near the center of the lowest Landau level
$\epsilon=0$. The three curves correspond to the following temperatures:
$T=0$ (solid line), $T=0.15\times\epsilon_\omega$
(broken line),
and $T=0.3\times\epsilon_\omega$ (dotted line). 
}
\end{figure}
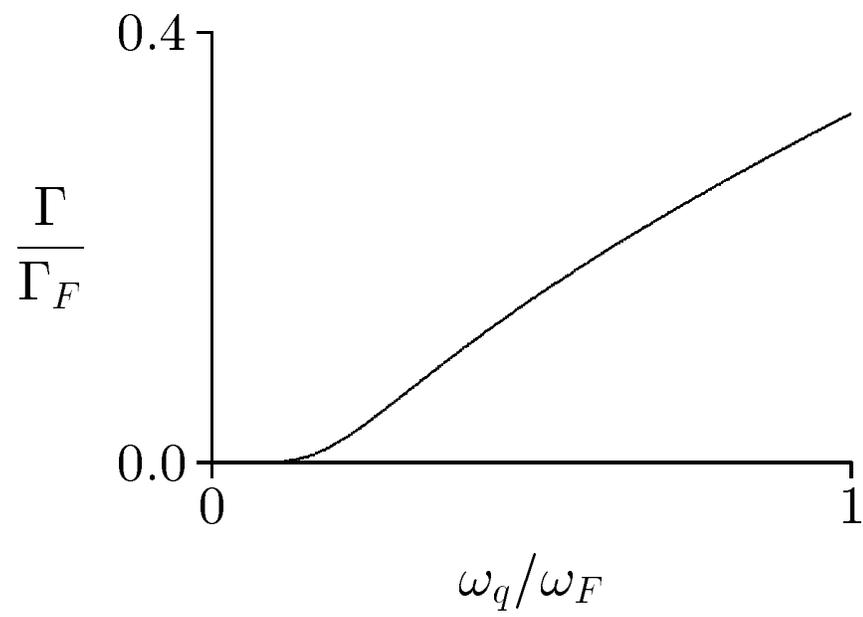
\begin{figure}
\caption{
The attenuation coefficient $\Gamma$, Eq.\ (\protect\ref{gscr}), as a function
of the SAW frequency $\omega_q$ for a fixed Fermi energy.
}
\end{figure}

\end{document}